\documentclass[preprint,amsmath,amssymb,aps,prd]{revtex4-2}
\usepackage{graphicx}
\usepackage{dcolumn}
\usepackage{bm}
\usepackage{color}
\begin{document}
%%%%%%%%%%%%%%%%%%%%%%%%%%%%%%%%%%%%%%%%%%%%%%%%%%%%%%%%%%%%%%%%%
\author{Gabriel \'Alvarez}
\author{Luis Mart\'{\i}nez Alonso}
\affiliation{Departamento de F\'{\i}sica Te\'orica,
                        Facultad de Ciencias F\'{\i}sicas,
                        Universidad Complutense,
                        28040 Madrid, Spain}
\author{Elena Medina}
\affiliation{Departamento de Matem\'aticas,
                        Facultad de Ciencias,
                        Universidad de C\'adiz,
                        11510 Puerto Real, C\'adiz, Spain}
%%%%%%%%%%%%%%%%%%%%%%%%%%%%%%%%%%%%%%%%%%%%%%%%%%%%%%%%%%%%%%%%%
\title{Kinetic dominance and the wavefunction of the universe}
%%%%%%%%%%%%%%%%%%%%%%%%%%%%%%%%%%%%%%%%%%%%%%%%%%%%%%%%%%%%%%%%%
\date{\today}
%%%%%%%%%%%%%%%%%%%%%%%%%%%%%%%%%%%%%%%%%%%%%%%%%%%%%%%%%%%%%%%%%
\begin{abstract}
We analyze the emergence of classical inflationary universes in a kinetic-dominated stage using a suitable class of
solutions of the Wheeler-De Witt equation with a constant potential. These solutions are eigenfunctions of the inflaton
momentum operator that are strongly peaked on classical solutions exhibiting either or both a kinetic dominated period
and an inflation period. Our analysis is based on semiclassical WKB solutions of the Wheeler-De Witt equation interpreted in the sense
of Borel (to perform a correct connection between classically allowed regions) and on the relationship
of these solutions to the solutions of the classical model. For large values of the scale factor the WKB
Vilenkin tunneling wavefunction and the Hartle-Hawking  no-boundary wavefunctions are recovered as particular
instances of  our class of wavefunctions.
\end{abstract}
%%%%%%%%%%%%%%%%%%%%%%%%%%%%%%%%%%%%%%%%%%%%%%%%%%%%%%%%%%%%%%%%%
\maketitle
%%%%%%%%%%%%%%%%%%%%%%%%%%%%%%%%%%%%%%%%%%%%%%%%%%%%%%%%%%%%%%%%%
\section{INTRODUCTION}\label{intro}
%%%%%%%%%%%%%%%%%%%%%%%%%%%%%%%%%%%%%%%%%%%%%%%%%%%%%%%%%%%%%%%%%
Inflationary cosmology is a successful theoretical framework  to understand the evolution and structure of
our Universe~\cite{STA80,GU81,LI82,LI85,MU05,BA09,MA18}. In particular, the study  of  the  appropriate
initial conditions for the emergence of an inflationary universe requires the use of quantum cosmology.
Thus, in the standard  descriptions of the early Universe (like those based on the Hartle-Hawking
no-boundary wavefunction~\cite{HH83,H84} or on the Vilenkin tunneling
wavefunction~\cite{VI86,VI88,Vi94,At94,LI08}), the interest is focused  on the Universe nucleation
at the appropriate  initial condition for inflation.

Nevertheless, a series of recent results~\cite{HAN14,HAN19,HAD18,HER19} (see also~\cite{ME20})
show that for broad classes of solutions of classical inflationary (INF) models the resulting early Universe
is in a state of \emph{kinetic dominance} (KD).  The KD period of inflationary models is a  pre-inflationary
stage~\cite{MU05,STEIN84,STE93,LID94,LIDS97,BA06,LI08,WE08,BO09,LA05,DE10,HAN14,HAN19}
which occurs when  the kinetic energy of the inflaton field dominates  over its potential energy,
a stage that quickly evolves to the INF stage. Moreover, it has been proved~\cite{HAN19} that KD
initial conditions are a  consistent alternative to the standard  INF, or slow-roll, initial conditions.

Moreover, due to the fact that during the  KD stage the comoving Hubble horizon grows, it follows that
initial KD conditions give rise to oscillations and to a cutof at large scales in the cosmic microwave  angular power spectra. 

Periods of KD behavior occur near a singularity $a=0$, and since a neighborhood of this singularity
is outside the range of validity  of the classical treatment, it is natural to expect that the transition
to the quantum  regime of INF models  might be  of a  KD character and that it might be studied with semiclassical approximations.
In the present work we analyze the  emergence of KD classical universes
using semiclassical  solutions of the Wheeler-De Witt (WDW) equation~\cite{DE67}.

More concretely, we consider single inflaton models  in a closed Friedmann universe
with unit curvature~\cite{HH83,H84,HP86,VI86,VI88,Vi94,At94}. The standard analysis of the
well-known Hartle-Hawking no-boundary~\cite{HH83,H84} and Vilenkin tunneling~\cite{VI86,VI88}
semiclassical wavefunctions does not lead to classically allowed regions of KD type, because these
wavefunctions are independent of the inflaton field for  small values of the scale factor. Some semiclassical
solutions of the WDW equation whose  corresponding classical solutions exhibit KD behavior have indeed
been considered in  Refs.~\cite{VI86,VI88}, but we could not find in the literature a more extensive
exploration of these solutions.

The usual hypothesis on the inflaton potential $V(\phi)$  of the WDW equation is that it is slowly varying
on a region  around a certain value $\phi_0$ of the inflaton~\cite{VI88, ha09}. In the present work we consider
wavefunctions that are eigenfunctions of the momentum conjugate to the inflaton variable, and show that under
the assumption of a constant inflaton potential $V(\phi_0)$,  the set of these wavefunctions
include a variety of families with WKB approximations  strongly peaked about  classical solutions on either or
both KD and INF states. Our discussion is guided by the properties of the solutions of the classical version
of the inflationary model with a constant potential and establishes a close correspondence between
its solutions and appropriate  wavefunctions of the WDW equation.  Moreover, our family of wavefunctions
includes  functions which for large values of the scale factor  $a$ reduce to products of  phase factor $\exp(ik\phi/\hbar)$
times Hartle-Hawking  no-boundary wavefunctions~\cite{HH83,H84} and Vilenkin tunneling wavefunctions~\cite{VI86,VI88},
while for $a$ near zero they manifest classically allowed regions of KD type.
We formulate an approximation scheme  to obtain  the explicit form of the  associated  WKB wavefunctions
and we obtain the corresponding  connection rules between classically allowed regions $\mbox{INF}\rightarrow\mbox{KD}$
and $\mbox{KD}\rightarrow\mbox{INF}$.  We also discuss the emergence of classical universes
in KD and  INF periods in terms of the solutions of the classical version of the model. We also consider the probability
distributions provided by  the  wavefunctions of the model representing expanding universes  with an inflation period.
Then we prove that the WKB approximations for   $a$ near zero  (KD region) and for large $a$ (INF region) lead to the same result.

The paper is organized as follows.  In Section~II  we describe the basic aspects of  the KD period of solutions
of classical inflationary models, and  in particular we introduce the solutions of the classical inflaton model with a constant inflaton
potential and analyze their KD and INF periods. In Sec.~III we review the properties of the WDW equation and of its WKB solutions,
paying special attention to the concept of WKB solutions peaked about classical solutions in classically allowed regions.
In Sec.~IV  we discuss the WDW equation with a constant potential. We introduce our family of
wavefunctions and determine the explicit form of their  associated  WKB wavefunctions.
We also obtain the corresponding WKB  connection rules. Section~V is devoted to the classification of emergent
universes which arise from our wavefunctons, and in Sec.~VI we discuss probability distributions.
Finally, we defer to an Appendix the exact solution of the WDW near $a=0$ that provides expressions
of the corresponding WKB wavefunctions on the KD region in terms of first-kind modified Bessel functions.
%%%%%%%%%%%%%%%%%%%%%%%%%%%%%%%%%%%%%%%%%%%%%%%%%%%%%%%%%%%%
\section{CLASSICAL INFLATON MODELS}\label{hj}
%%%%%%%%%%%%%%%%%%%%%%%%%%%%%%%%%%%%%%%%%%%%%%%%%%%%%%%%%%%%
We  consider classical single-field  inflaton models in a closed Friedmann universe with unit
curvature~\cite{LI85,MU05,BA09,MA18},
\begin{equation}
	\label{eq0}
	d s^2=-d t^2+a(t)^2 \left[ \frac{d r^2}{1- r^2}+r^2(d \theta^2+\sin^2 \theta d \phi^2) \right].
\end{equation}
The dynamical variables are the scale factor $a(t)\geq 0$ and  the real field $-\infty<\phi(t)<\infty$,
which are functions of the cosmic time $t$. They satisfy the evolution equation,
\begin{equation}
	\label{eq1}
	\ddot{\phi} + 3H\dot{\phi} + V'(\phi) = 0,
\end{equation}
and the Friedmann equation,
\begin{equation}
	\label{eq2}
	H^2 = \frac{1}{3M_{\rm pl}^2}\left(\frac{1}{2}\dot{\phi}^2+V(\phi)\right)-\frac{1}{a^2},
\end{equation}
where  dots denote derivatives with respect to $t$, $H=\dot{a}/a$ is the Hubble parameter, $V(\phi)$
is a given positive smooth potential, and $M_{\rm pl}=\sqrt{\hbar c/8\pi G}$ is the reduced Planck mass.

The energy density is
\begin{equation}
	\label{den}
	\rho = \frac{1}{2}\dot{\phi}^2+V(\phi),
\end{equation}
which using Eq.~(\ref{eq2}) can be expressed as
\begin{equation}
	\label{den1}
	\rho = 3 M_{\rm pl}^2 \left( H^2+\frac{1}{a^2} \right),
\end{equation}
while its cosmic-time derivative, which follows from Eqs.~(\ref{eq1}) and~(\ref{eq2}), is
\begin{equation}
	\label{den3}
	\dot{\rho} = -3 H \dot{\phi}^2.
\end{equation}

Note also that for the classical inflaton models to be reliable, the energy density $\rho$ must be smaller
than the Planck density $\rho_{\rm pl}=M_{\rm pl}^4$, i.e.,
\begin{equation}
	\label{den2}
	\rho <  M_{\rm pl}^4,
\end{equation}
which combined with Eq.~(\ref{den1}) give a  classical lower limit for  the scale factor
\begin{equation}
	\label{bo}
	a > \frac{\sqrt{3}}{M_{\rm pl}}.
\end{equation}
%%%%%%%%%%%%%%%%%%%%%%%%%%%%%%%%%%%%%%%%%%%%%%%%%%%%%%%%%%%%
\subsection{Inflationary regime and kinetic dominance}
%%%%%%%%%%%%%%%%%%%%%%%%%%%%%%%%%%%%%%%%%%%%%%%%%%%%%%%%%%%%
The acceleration equation,
\begin{equation}
	\label{in1}
	\frac{\ddot{a}}{a} = -\frac{1}{3M_{\rm pl}^2} ( \dot{\phi}^2-V(\phi) ),
\end{equation}
which also follows from Eqs.~(\ref{eq1}) and~(\ref{eq2}), shows that the inflationary regime $\ddot{a}>0$
is determined by the condition
\begin{equation}
	\dot{\phi}^2 < V(\phi),
\end{equation}
or, in terms of the energy density,
\begin{equation}
	\label{in 2}
	\rho < \frac{3}{2} V.
\end{equation}
Therefore, as a consequence of Eq.~(\ref{den1}),  during the inflationary period
 \begin{equation}
 	\label{inf}
	 a > \sqrt{\frac{2}{V}}M_{\rm pl}.
 \end{equation}

Kinetic dominance is the opposite regime, wherein the energy density is dominated by the kinetic energy of the inflaton field,
\begin{equation}
	\label{kd0}
	\dot{\phi}^2 \gg V(\phi),
\end{equation}
or, equivalently,
\begin{equation}
	\label{kd}
	\rho \gg \frac{3}{2} V.
\end{equation}
In the KD regime we may neglect $V$,  $V'$ and $1/a^2$  in Eqs.~(\ref{eq1}) and~(\ref{eq2}),
which then decouple into an equation for the inflaton field,
\begin{equation}
    \label{eq1a}
    \ddot{\phi} \pm \sqrt{\frac{3}{2}}\frac{1}{{M_{\rm pl}}} |\dot{\phi}|\,\dot{\phi}\sim 0,
\end{equation}
and an equation to calculate $H = \dot{a}/a$ given any solution $\phi(t)$ of the former,
\begin{equation}
	\label{eq1b}
	 \frac{\dot{a}}{a} \sim \pm  \frac{1}{\sqrt{6} \,{M_{\rm pl}}}|\dot{\phi}  |.
\end{equation}
Thus we obtain the following two families of  asymptotic solutions of Eqs.~(\ref{eq1})--(\ref{eq2}) near a singularity $a=0$:
for the plus sign in Eq.~(\ref{eq1a}), solutions expanding from the singularity,
\begin{equation}
	\label{phiasy}
	\phi(t)\sim \pm\sqrt{\frac{2}{3}}{M_{\rm pl}}\log(t-t^*)+\phi^*,
	\quad a(t) \sim a^* (t-t^*)^{1/3},
	\mbox{ as }t\rightarrow(t^*)^+,
\end{equation}
and for the minus sign in Eq.~(\ref{eq1a}), solutions collapsing towards the singularity,
\begin{equation}
	\label{phiasy2}
	\phi(t)\sim \pm\sqrt{\frac{2}{3}}{M_{\rm pl}}\log(t^*-t)+\phi^*,
	\quad a(t) \sim a^* (t^*-t)^{1/3},
	\mbox{ as } t\rightarrow(t^*)^-,
\end{equation}
where $t^*$, $\phi^*$  and  $a^*>0$ are arbitrary integration constants.  Note also that these asymptotic solutions are
independent of the inflaton potential $V(\phi)$. This behavior is illustrated
in Fig.~1, which shows the results of numerical integrations of Eqs.~(\ref{eq1})--(\ref{eq2}) for the quadratic potential
$V(\phi)=m^2 \phi^2$ and for the Starobinski potential  $V(\phi)=\Lambda^2[1-\exp(-\sqrt{2/3}\phi/{M_{\rm pl}})]^2$.
(The KD regime may have different behaviors if the potential is not everywhere nonnegative~\cite{FE02}.
We do not consider those potentials in this paper). 
%%%%%%%%%%%%%%%%%%%%%%%%%%%%%%%%%%%%%%%%%%%%%%%%%%%%%%%%%%%%
\begin{figure}
\begin{center}
        \includegraphics[width=7cm]{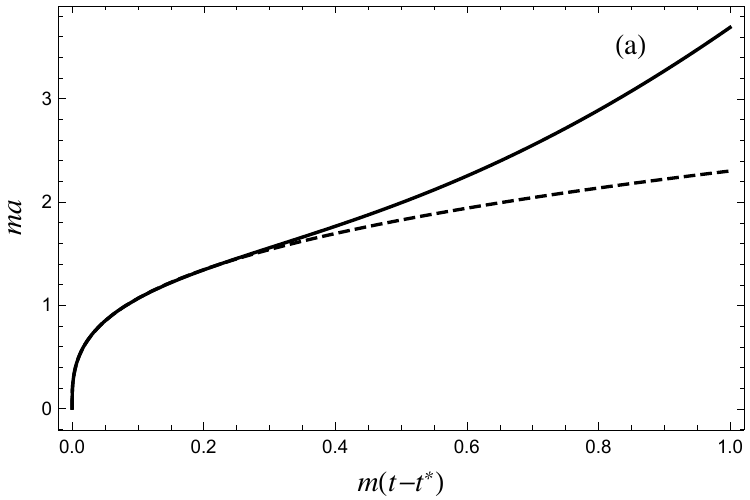}
        \hspace{1cm}
        \includegraphics[width=7cm]{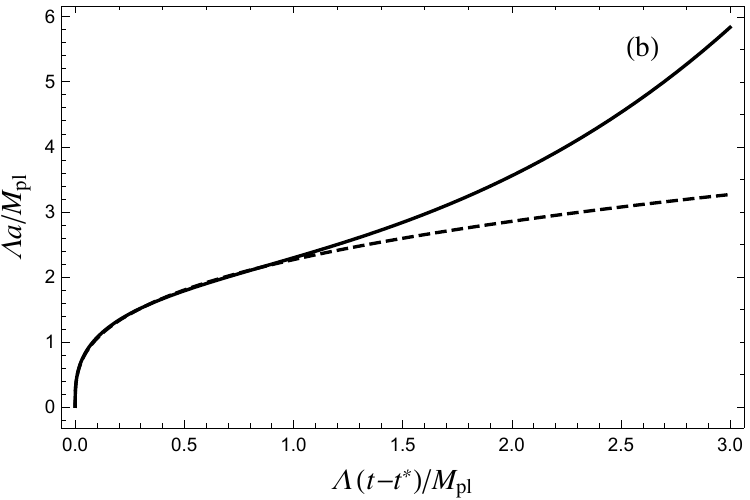}
\end{center}
   \caption{
    Scale factor as a function of cosmic time obtained by numerical integration of  Eqs.~(\ref{eq1})--(\ref{eq2})
    (continuous lines) for:
    (a) the quadratic potential  $V(\phi)=m^2 \phi^2$ with initial values
         $t_0=0$, $\phi_0=M_{\rm pl}$, $\dot{\phi}_0 = 10 m M_{\rm pl}$, and $a_0 = m^{-1}$;
    (b) the Starobinski  potential  $V(\phi)=\Lambda^2[1-\exp(-\sqrt{2/3}\phi/{M_{\rm pl}})]^2$ with initial values
         $t_0=0$, $\phi_0=M_{\rm pl}$, $\dot{\phi}_0 = 10 \Lambda$, $a_0 = M_{\rm pl}/\Lambda$.
    Both models exhibit the KD behavior (dashed lines) $a(t)\sim a^* (t-t^*)^{1/3}$  as $t\rightarrow(t^*)^+$.}
\end{figure}
%%%%%%%%%%%%%%%%%%%%%%%%%%%%%%%%%%%%%%%%%%%%%%%%%%%%%%%%%%%%%%%%%
%%%%%%%%%%%%%%%%%%%%%%%%%%%%%%%%%%%%%%%%%%%%%%%%%%%%%%%%%%%%%%%%%
\subsection{ The classical inflaton model with a constant potential.}
%%%%%%%%%%%%%%%%%%%%%%%%%%%%%%%%%%%%%%%%%%%%%%%%%%%%%%%%%%%%%%%%%
%%%%%%%%%%%%%%%%%%%%%%%%%%%%%%%%%%%%%%%%%%%%%%%%%%%%%%%%%%%%%%%%%
\begin{figure}
\begin{center}
        \includegraphics[width=14cm]{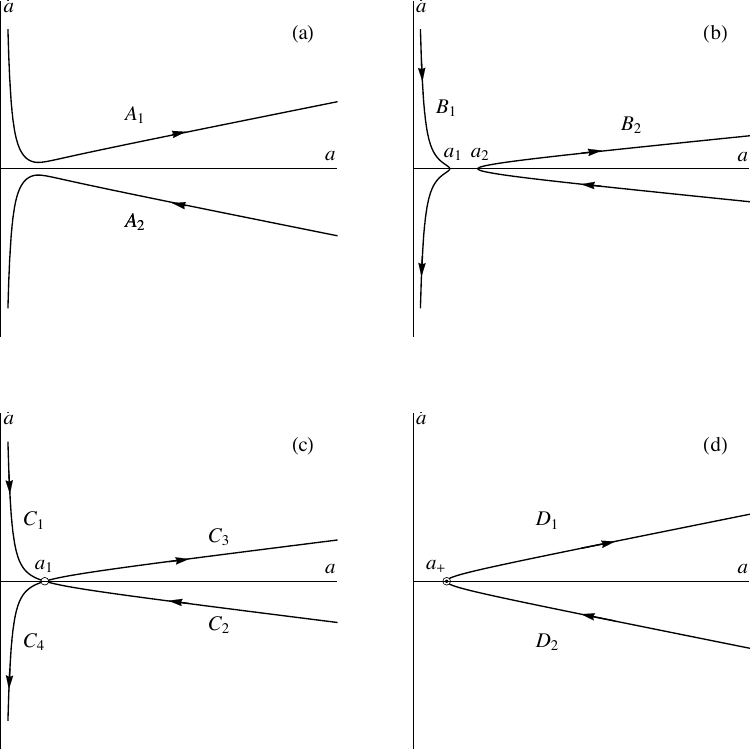}
      \end{center}

         \caption{Trajectories in the $(a,\dot{a})$ phase space of the solutions of the classical model with constant inflaton potential
                 given by Eq.~(\ref{co2}) in the four possible cases:
                 (a) $(4/27)a_+^4 < b_+$; (b) $0<b_+< (4/27) a_+^4$; (c) $b_+ = (4/27) a_+^4$; and (d) $b_+= 0$.}
\end{figure}
%%%%%%%%%%%%%%%%%%%%%%%%%%%%%%%%%%%%%%%%%%%%%%%%%%%%%%%%%%%%%%%%%

We consider inflaton models with a constant potential as their solutions  exhibit  one or both periods of kinetic dominance
and of inflationary expansion. These models are relevant on  the domain of interest of models with slowly-varying potentials
around a value $\phi_0$, where  the potential  may be approximated by a constant $V_0$,
\begin{equation}
	\label{vco}
	V(\phi) \approx V(\phi_0) = V_0,
\end{equation}
For example, the  Starobinski  model with  potential  $V(\phi)=\Lambda^2[1-\exp(-\sqrt{2/3}\phi/{M_{\rm pl}})]^2$
is slowly-varying around any large value of $\phi$, where $V(\phi) \approx \Lambda^2$.

For a constant potential Eq~(\ref{eq1}) reduces to
 \begin{equation}
 	\label{co1}
 	\dot{\phi}(t) a(t)^3 = \dot{\phi}(t_0)a(t_0)^3,
 \end{equation}
where $t_0$ is a given value of the cosmic time, and  Eq.~(\ref{eq2}) can be written as
 \begin{equation}
 	\label{co2}
 	\dot{a}^2 = \frac{a^2}{a_+^2}+\frac{b_+}{a^4}-1,
 \end{equation}
 where
 \begin{equation}
 	\label{co3}
 	a_+ = M_{\rm pl} \sqrt{3/V_0},
 \end{equation}
and
 \begin{equation}
 	\label{co4}
 	b_+ = \frac{(\dot{\phi}(t_0)a(t_0)^3)^2}{6{M_{\rm pl}}^2}.
 \end{equation}

Equation~(\ref{co2})  can be solved by separation of variables,
 \begin{equation}
 	\label{integ}
 	t-t_0 = \pm\int_{a(t_0)}^{a(t)} \frac{d a}{\sqrt{\frac{a^2}{a_+^2}+\frac{b_+}{a^4}-1}}.
 \end{equation}
These solutions can be classified into four types depending on the number of zeros of the
right-hand side of Eq.~(\ref{co2}) on the domain $a>0$. For convenience, in the following
discussion we denote this right-hand side by $f(a)$, i.e., $f(a) = a^2/a_+^2+b_+/a^4-1$.
%%%%%%%%%%%%%%%%%%%%%%%%%%%%%%%%%%%%%%%%%%%%%%%%%%%%%%%%%%%%%%%%%
\begin{itemize}
	\item[(a)]	If  $(4/27)a_+^4 < b_+$, then there are no zeros of $f(a)$ in the domain $a>0$, and the solutions give rise to the two
			families of trajectories in the $(a,\dot{a})$ phase map shown in Fig.~2(a).
			The upper trajectory $A_1$ describes an expanding universe starting from a KD stage near a singularity $a=0$ at
			cosmic time
			 \begin{equation}
			 	\label{co4+}
 				t^* = t_0+\int_{a(t_0)}^{0} \frac{d a}{\sqrt{\frac{a^2}{a_+^2}+\frac{b_+}{a^4}-1}},
 			\end{equation}
			such that  $\dot{a}$ first decreases towards a positive minimum value at
			\begin{equation}
				\label{ain}
				a_{\text{in}} = (2b_+)^{1/6}\, a_+^{1/3},
			\end{equation}
			and then enters an unlimited period of inflationary expansion. The solutions giving rise to the trajectory
			$A_2$ describe the corresponding contracting, time-reversed evolution.

	\item[(b)] If  $0<b_+< (4/27) a_+^4$, then $f(a)$ has two positive zeros $0<a_1<a_2$,
			\begin{equation}
				\label{solap0}
				a_k = a_+\sqrt{\frac{1}{3} +
				\frac{2}{3}\cos\left( \frac{1}{3}\arccos\left(1-\frac{27\,b_+}{2\,a_+^4}\right)-(k-1)\frac{2\pi}{3}\right)},
				\quad k=1,2,
			\end{equation}
			and there are the two classes of trajectories shown in Fig.~2(b).
			The trajectory $B_1$ represents solutions with $a\leq a_1$ that describe universes which
			expand from a KD stage near a singularity $a=0$ until they reach a maximum $a(t)=a_1$,
			and then contract towards another KD stage near $a=0$. These solutions do not have an INF period.
			The trajectory  $B_2$ represents solutions with $a\geq a_2$  which describe universes without any KD period:
			they contract until they bounce at $a=a_2$ and begin a period of unlimited INF expansion.

	\item[(c)] If $b_+ = (4/27) a_+^4$, $f(a)$ has a unique positive zero at
    			\begin{equation}
				\label{consts}
  				a_0 = \sqrt{\frac{2}{3}}\, a_+,
			\end{equation}
 			but $f(a)\geq 0$ for $a>0$.
			The corresponding trajectories are shown in Fig.~2(c), which includes the constant solution $a=a_0$
  			as well as four families of trajectories, two of which ($C_1$ and $C_2$) tend asymptotically to $a_0$
			in the future, and the other two ($C_3$ and $C_4$) in the past.

	\item[(d)] Finally, if $b_+= 0$, then $f(a)$ has a unique positive zero at $a=a_+$, but $f(a)\geq 0$ only for $a\geq a_+$.
    			The corresponding trajectories are shown in Fig.~2(d),  which features the constant solutions $a=a_+$
  			as well as two  families of solutions ($D_1$, expanding, and $D_2$, contracting),  without KD periods,
			and which  tend asymptotically to $a_+$  in the past and in the future, respectively.
			The solutions corresponding to $D_1$ are in an unlimited period of INF expansion.
			The particular case $a(t_0)=a_+$ can be explicitly integrated to give the de~Sitter form,
			\begin{equation}\label{inso}
  				a(t) = a_+\cosh\left(\frac{1}{a_+}(t-t_0)\right).
  			\end{equation}
\end{itemize}
%%%%%%%%%%%%%%%%%%%%%%%%%%%%%%%%%%%%%%%%%%%%%%%%%%%%%%%%%%%%%%%%%
\section{The  WDW equation and its WKB solutions }
%%%%%%%%%%%%%%%%%%%%%%%%%%%%%%%%%%%%%%%%%%%%%%%%%%%%%%%%%%%%%%%%%
In this Section we briefly recall the derivation of the WDW equation, the structure of its WKB solutions, and the relation of the
latter to the classical inflaton Eqs.~(\ref{eq1})--(\ref{eq2}). This relation will be particularized in the following Section to
the case of a constant inflaton potential, which in turn will be used to discuss possible types of emerging universes.
%%%%%%%%%%%%%%%%%%%%%%%%%%%%%%%%%%%%%%%%%%%%%%%%%%%%%%%%%%%%%%%%%
%%%%%%%%%%%%%%%%%%%%%%%%%%%%%%%%%%%%%%%%%%%%%%%%%%%%%%%%%%%%%%%%%
\subsection{The WDW equation}
%%%%%%%%%%%%%%%%%%%%%%%%%%%%%%%%%%%%%%%%%%%%%%%%%%%%%%%%%%%%%%%%%
The momenta conjugate to the variables $a$ and $\phi$ in the Hamiltonian formulation of the inflaton model turn out
to be~\cite{VI88}
\begin{equation}
	\label{moa}
	P_a=-12 \pi^2{M_{\rm pl}}^2 a \dot{a},
\end{equation}
and
\begin{equation}
	\label{mof}
	P_{\phi}=2\pi^2 a^3 \dot{\phi},
\end{equation}
respectively. If we substitute these values in Eq.~(\ref{eq2}) rewritten in the form
\begin{equation}
	\label{eq2b}
	6{M_{\rm pl}}^2 a\,\dot{a}^2- a^3\, \dot{\phi}^2+6{M_{\rm pl}}^2\,  \left(a-\frac{V(\phi)}{3{M_{\rm pl}}^2}a^3\right)=0,
\end{equation}
we arrive at
\begin{equation}
	\label{eq2c}
	P_a^2 - \frac{6 {M_{\rm pl}}^2}{a^2}\, P_{\phi}^2 + U(a,\phi)=0,
\end{equation}
where $U(a,\phi)$ is the \emph{superpotential}  function
\begin{equation}
	\label{par}
	U(a,\phi)=\frac{1}{\lambda^2}\left(a^2- \frac{V(\phi)}{3{M_{\rm pl}}^2}a^4\right),
\end{equation}
and where
\begin{equation}
	\label{pa2r}
	\lambda = \frac{1}{12 \pi^2 {M_{\rm pl}}^2}.
\end{equation}
Incidentally, since the left-hand side of Eq.~(\ref{eq2c}) is proportional to $a\mathcal{H}$,
where $\mathcal{H}$ is the Hamiltonian of the inflaton model~\cite{VI88}, Eq.~(\ref{eq2c}) shows that the solutions
of Eqs.~(\ref{eq1})--(\ref{eq2}) satisfy the zero-energy constraint $\mathcal{H}=0$.

Quantization is performed by the rules,
\begin{equation}
	\label{ord}
	P_a \rightarrow -i\, \hbar \frac{\partial}{\partial a}, \quad P_{\phi} \rightarrow -i\, \hbar \frac{\partial}{\partial \phi},
\end{equation}
with the proviso of an ambiguity in the ordering of $a$ and $\partial /\partial a$ in the quantization of $P_a^2$,
which in effect leads to the quantization rule
\begin{equation}
	\label{ord1}
	P_a^2 \rightarrow -\hbar^2\frac{1}{a^p} \frac{\partial}{\partial a} a^p  \frac{\partial}{\partial a},
\end{equation}
where $p$ is an (in principle) arbitrary parameter. Thus we arrive at the celebrated WDW equation~\cite{DE67},
\begin{equation}
	\label{WDW}
	\left(
		\frac{\partial^2}{\partial a^2}+\frac{p}{a} \frac{\partial }{\partial a}
		-
		\frac{6 {M_{\rm pl}}^2}{a^2} \frac{\partial^2}{\partial \phi^2}-\frac{1}{\hbar^{2}}U(a,\phi)
	\right)\Psi(a,\phi)=0.
\end{equation}
%%%%%%%%%%%%%%%%%%%%%%%%%%%%%%%%%%%%%%%%%%%%%%%%%%%%%%%%%%%%%%%%%
\subsection{General remarks on the WKB solutions of the WDW equation}
%%%%%%%%%%%%%%%%%%%%%%%%%%%%%%%%%%%%%%%%%%%%%%%%%%%%%%%%%%%%%%%%%
WKB solutions of Eq.~(\ref{WDW}) have the usual exponential form
\begin{equation}
	\label{oss1}
	\Psi(a,\phi) = \exp\left( i \,\mathbb{S}(a,\phi)/\hbar \right),
\end{equation}
where $\mathbb{S}$ is expanded as an asymptotic power series in $\hbar$,
\begin{equation}
	\label{oss2}
	\mathbb{S}=  \mathbb{S}_0+\frac{\hbar}{i} \mathbb{S}_1+\left(\frac{\hbar}{i}\right)^2 \mathbb{S}_2+\cdots ,\quad \hbar\rightarrow 0.
\end{equation}
By substituting Eqs.~(\ref{oss1})--(\ref{oss2}) into the WDW Eq.~(\ref{WDW}) and setting to zero terms with the same power of $\hbar$
we obtain a sequence of two-dimensional WKB equations for the $\mathbb{S}_n$, which have to be solved recursively.

The first equation, for $\mathbb{S}_0$, is the Hamilton-Jacobi equation,
\begin{equation}
	\label{firep}
	\left(\frac{\partial  \mathbb{S}_0}{ \partial a}\right)^2
	-\frac{1}{2\pi^2\lambda a^2}\left(\frac{\partial \mathbb{S}_0}{ \partial \phi}\right)^2
	+U=0,
\end{equation}
and the second equation, for $\mathbb{S}_1$, is
\begin{equation}
	\label{nex}
	\frac{\partial ^2 \mathbb{S}_0}{ \partial a^2}
	-\frac{6 {M_{\rm pl}}^2}{a^2}\frac{\partial ^2 \mathbb{S}_0}{ \partial \phi^2}
	+2\frac{\partial  \mathbb{S}_0}{ \partial a}\frac{\partial  \mathbb{S}_1}{ \partial a}
	-\frac{12 {M_{\rm pl}}^2}{a^2}\frac{\partial  \mathbb{S}_0}{ \partial \phi}\frac{\partial  \mathbb{S}_1}{ \partial \phi}
	+\frac{p}{a}\frac{\partial  \mathbb{S}_0}{ \partial a}=0.
\end{equation}
Note that the Hamilton-Jacobi Eq.~(\ref{firep}) does not depend on the ordering parameter $p$, while Eq.~(\ref{nex}) does.

WKB solutions of the WDW equation in the classically allowed regions are oscillatory, and these are the regions where
the nucleation of classical universes with specific properties may emerge~\cite{At94,ha09}. We also recall that the
potential $V(\phi)$ must be positive and smaller than the Planck density ${M_{\rm pl}}^4$ for the
semiclassical solutions obtained from Eqs.~(\ref{firep})--(\ref{nex}) to be valid~\cite{VI86}.
The identification between the classical momenta as defined in Eqs.~(\ref{moa})--(\ref{mof}) and
the corresponding derivatives of $\mathbb{S}_0$,
\begin{equation}
	\label{idee}
	P_a=\frac{\partial  \mathbb{S}_0}{\partial a},\quad P_{\phi}=\frac{\partial  \mathbb{S}_0}{\partial \phi},
\end{equation}
leads to the system of first-order differential equations,
\begin{equation}
	\label{sysc}
	\dot{a} = -\frac{\lambda}{a}\frac{\partial  \mathbb{S}_0}{\partial a},
	\quad
	\dot{\phi}=\frac{1}{2\pi^2 a^3}\frac{\partial  \mathbb{S}_0}{\partial \phi}.
\end{equation}
Thus, each solution of Eq.~(\ref{firep}) corresponds to a biparametric (the initial conditions for $a$ and $\phi$)
family of solutions of Eqs.~(\ref{eq1})--(\ref{eq2}), and this family in turn characterizes the properties of the emergent classical
universes in the classically allowed regions. It is then said~\cite{At94,ha09} that the wavefunction
$\Psi(a,\phi)$ is \emph{peaked} about the associated solutions of the
system~(\ref{sysc}).
%%%%%%%%%%%%%%%%%%%%%%%%%%%%%%%%%%%%%%%%%%%%%%%%%%%%%%%%%%%%%%%%%
\section{THE WDW EQUATION WITH A CONSTANT POTENTIAL}
%%%%%%%%%%%%%%%%%%%%%%%%%%%%%%%%%%%%%%%%%%%%%%%%%%%%%%%%%%%%%%%%%
In this Section we look for solutions of the WDW equation with a constant potential using the ansatz
\begin{equation}
	\label{nac}
	\Psi(k,a,\phi) = \psi(k,a) e^{i k \phi/\hbar},
\end{equation}
where $k$ is a real parameter. Variables in the WDW equation separate and we get the following equation for $\psi(k,a)$,
\begin{equation}
	\label{WDW1}
	\left(  \frac{d^2}{d a^2}+\frac{p}{a}\frac{d}{d a}-\hbar^{-2}\mathrm{U}(k,a) \right) \psi(k,a) = 0,
\end{equation}
where
\begin{equation}
	\label{pa2rrr}
	\mathrm{U}(k,a) = -\frac{6{M_{\rm pl}}^2 k^2}{a^2}+U(a,\phi_0)
	                          = -\frac{6{M_{\rm pl}}^2 k^2}{a^2}+ \frac{a^2}{\lambda^2} \left(1-\frac{a^2}{ a_+^2}\right).
\end{equation}
The first term in the modified superpotential $\mathrm{U}(k,a)$ is a consequence of the phase factor
$\exp(i k \phi/\hbar)$ in the wavefunction ansatz Eq.~(\ref{nac}), and induces a dependence of the solutions
of the WDW equation on the scale factor different from those in Refs.~\cite{VI88,ha09}. This term is dominant
as $a\to 0^+$, and in the next Sections we will discuss its effect on the overall behavior of the WKB solutions
via connection formulas.
%%%%%%%%%%%%%%%%%%%%%%%%%%%%%%%%%%%%%%%%%%%%%%%%%%%%%%%%%%%%%%%%%
\subsection{WKB solutions}
%%%%%%%%%%%%%%%%%%%%%%%%%%%%%%%%%%%%%%%%%%%%%%%%%%%%%%%%%%%%%%%%%
The actions $\mathbb{S}$ and $S$ of the WKB solutions for $\Psi(k,a,\phi)$ and
\begin{equation}
	\psi(k,a) = e^{i S(k,a)/\hbar},
\end{equation}
in Eq.~(\ref{nac}) are related by $\mathbb{S}=S+k\phi$, which after expansion in powers of $\hbar$ leads to
$\mathbb{S}_0 = S_0 + k\phi$ and $\mathbb{S}_n = S_n$ for $n\geq 1$ (with $S_n$ independent of $\phi$).
Therefore, the classical system Eq.~(\ref{sysc}) on whose solutions the wavefunctions $\Psi(k,a,\phi)$ are strongly peaked is,
\begin{equation}
	\label{coo1}
	\dot{a}^2=-\frac{\lambda^2}{a^2}\mathrm{U}(k,a),\quad \dot{\phi}=\frac{k}{2\pi^2 a^3},
\end{equation}
which using Eq.~(\ref{pa2rrr}) reduces to  Eqs.~(\ref{co1})--(\ref{co2}) with
 \begin{equation}
 	\label{co3b}
 	b_+=\frac{k^2}{24 \pi^4 {M_{\rm pl}}^2}.
 \end{equation}
Likewise, Eqs.~(\ref{firep}) and~(\ref{nex}) reduce to equations for $S_0$ and $S_1$,
\begin{equation}
	\label{fir}
	\left(\frac{d S_0}{ d a}\right)^2 +\mathrm{U}(k,a)=0,
\end{equation}
\begin{equation}
	\label{fir3}
	\frac{d^2 S_0}{ d a^2}+2\frac{d S_0}{ d a}\frac{d S_1}{ d a}+\frac{p}{a}\frac{d S_0}{ d a}=0,
\end{equation}
which can be readily solved. On the classically allowed regions $\mathrm{U}(k,a)<0$ we find
\begin{equation}
	S_0(k,a) = \int_{a_r}^a \mathrm{p}(k,a') d a',
\end{equation}
where
\begin{equation}
	\label{mom}
	\mathrm{p}(k,a) = \left( -\mathrm{U}(k,a)\right)^{1/2},
\end{equation}
$a_r$ is an appropriate reference point, and
\begin{equation}
	\label{fir000}
	S_1(k,a) = -\frac{1}{2} \log(a^p \mathrm{p}(k,a)).
\end{equation}
Therefore, the basic WKB  solutions of Eq.~(\ref{WDW1}) on the classically allowed regions are
\begin{equation}
	\label{ca2}
	\psi_{\pm}(k,a)= \frac{1}{(a^p\mathrm{p}(k,a))^{1/2}} \exp \left( \pm \frac{i}{\hbar}\int_{a_r}^a \mathrm{p}(k,a') d a' \right).
\end{equation}

Eq.~(\ref{pa2rrr}) shows that for $k\neq 0$ the modified superpotential $\mathrm{U}(k,a)$ is strictly negative both for large
and small values of $a>0$, and therefore these regions are classically allowed regions for nonvanishing values of $k$,
values to which we restrict hereafter. The classical system Eq.~(\ref{sysc}) on whose solutions the wavefunctions $\psi$
are strongly peaked is
\begin{equation}
	\label{coo}
	\dot{a} = \mp \frac{\lambda}{a}\mathrm{p}(k,a),\quad \dot{\phi}=\frac{k}{2\pi^2 a^3},
\end{equation}
which reduces to Eqs.~(\ref{co1})--(\ref{co2}) with $b_+$ defined in Eq.~(\ref{co3b}). Note in particular that
the first equation of this system shows that $\psi_{+}$ and $\psi_{-}$ describe contracting and expanding universes,
respectively.
%%%%%%%%%%%%%%%%%%%%%%%%%%%%%%%%%%%%%%%%%%%%%%%%%%%%%%%%%%%%%%%%%
\subsection{Leading behavior of the WKB solutions at large scale factor}
%%%%%%%%%%%%%%%%%%%%%%%%%%%%%%%%%%%%%%%%%%%%%%%%%%%%%%%%%%%%%%%%%
As we mentioned earlier, at large $a$ we may neglect the $k$-dependent term in the modified superpotential Eq.~(\ref{pa2rrr}),
\begin{equation}
	\label{u+}
	\mathrm{U}(k,a) \sim U_+(a) = \frac{a^2}{\lambda^2}\left(1- \frac{a^2}{a_+^2}\right),
	\quad
	\mbox{ as } a\to\infty,
\end{equation}
in effect recovering the well-known model for studying the emergence of inflation from  the WDW equation
discussed, e.g., in Refs.~\cite{VI88, ha09},
\begin{equation}
	\label{WDW1+}
	\left( \frac{d^2}{d a^2}+\frac{p}{a}\frac{d}{d a}-\hbar^{-2}U_+(a) \right) \psi^{(\infty)}(a)=0.
\end{equation}
Since in our case $k\neq 0$ these results apply only to large $a$, we have labeled the wavefunction as $\psi^{(\infty)}(a)$.
We summarize here the relevant notations and results.

The generic shape of the superpotential $U_+(a)$ in shown in Fig.~3, and corresponds to a quantum
tunneling problem with a classically forbidden region $0< a < a_+$ and a classically allowed region $a > a_+$.
%%%%%%%%%%%%%%%%%%%%%%%%%%%%%%%%%%%%%%%%%%%%%%%%%%%%%%%%%%%%%%%%%
\begin{figure}
\begin{center}
        \includegraphics[width=8cm]{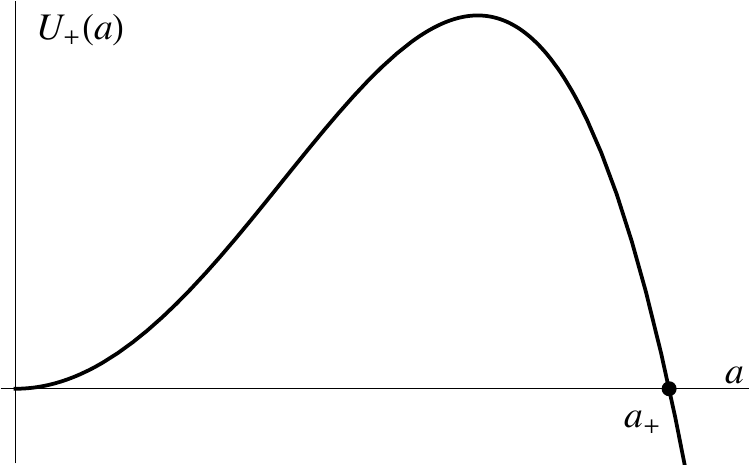}
 \end{center}
    \caption{Generic shape of the superpotential $U_+(a)$ defined in Eq.~(\ref{u+}).}
\end{figure}
%%%%%%%%%%%%%%%%%%%%%%%%%%%%%%%%%%%%%%%%%%%%%%%%%%%%%%%%%%%%%%%%%

Since $a_+$ is the unique positive zero of $U_+(a)$, it is natural to take it as the reference
point for the WKB solutions. By using Eq.~(\ref{ca2})  with $\mathrm{U}(k,a)$ replaced by $U_+(a)$
and $a_r=a_+$, we obtain the following ($k$-independent)  approximations $\psi_{\pm}^{(\infty)}(a)$
to the ($k$-dependent) functions $\psi_{\pm}(k,a)$ valid for large $a$,
\begin{equation}
	\label{wke}
	\psi_{\pm}^{(\infty)}(a)
	= \lambda^{1/2}a^{-p/2}\left(\frac{a^4}{a_+^2} -a^2\right)^{-1/4}
	\exp \left[\pm \frac{i}{\hbar} \,   \frac{{M_{\rm pl}}^2}{\lambda V_0}\left(\frac{a^2}{a_+^2} -1\right)^{3/2} \right],
	\quad
	a_+ < a.
\end{equation}
The wavefunctions $\psi_{+}^{(\infty)}(a)$ and $\psi_{-}^{(\infty)}(a)$ represent contracting and expanding universes in the inflationary period,
respectively.

Two distinguished wavefunctions are usually defined in terms of the $\psi_{\pm}^{(\infty)}(a)$.
The Vilenkin \emph{tunneling} wavefunction  $\psi_\mathrm{T}(a)$ is defined as~\cite{VI88, ha09},
\begin{equation}
 	\label{vil}
 	\psi_\mathrm{T}(a>a_+)= e^{ i\pi/4}A\psi_{-}^{(\infty)}(a),
\end{equation}
where
\begin{equation}
	\label{v444}
  	A = e^{-{M_{\rm pl}}^2/(\hbar \lambda V_0)}.
\end{equation}
The function $\psi_\mathrm{T}(a)$ represents an expanding universe in the region  $a>a_+$, and the corresponding
complete wavefunction in that region is
\begin{equation}\label{completeT}
	\Psi_\mathrm{T}(k,a,\phi) = \psi_\mathrm{T}(a) e^{i k\phi/\hbar}.
\end{equation}

Likewise, the Hartle-Hawking \emph{no-boundary} wavefunction~\cite{HH83,H84} is defined by the following
superposition of wavefunctions representing  contracting and expanding universes,
\begin{equation}
	\label{h3o0}
	\psi_\mathrm{NB}(a>a_+) =  A^{-1} ( e^{- i\pi/4}\psi_+^{(\infty)}(a)+e^{+ i\pi/4}\psi_-^{(\infty)}(a) ),
\end{equation}
and the corresponding complete wavefunction is
\begin{equation}
	\label{completeNB}
	\Psi_\mathrm{NB}(k,a,\phi) = \psi_\mathrm{NB}(a) e^{i k\phi/\hbar}.
\end{equation}

Note that if $V_0\ll {M_{\rm pl}}^4$, then $a_+$ is much larger than the classical lower limit for the scale factor
given by Eq.~(\ref{bo}),
 \begin{equation}
 	\label{des}
 	a_+ \gg \frac{\sqrt{3}}{{M_{\rm pl}}},
\end{equation}
and from Eq.~(\ref{inf}) it follows that during the classical inflation regime
\begin{equation}
	\label{isd}
	a >\sqrt{\frac{2}{3}}a_+.
\end{equation}
Therefore, $a_+$ is an appropriate lower limit for a classical inflationary stage.
%%%%%%%%%%%%%%%%%%%%%%%%%%%%%%%%%%%%%%%%%%%%%%%%%%%%%%%%%%%%%%%%%
\subsection{Leading behavior of the WKB solutions at small scale factor}
%%%%%%%%%%%%%%%%%%%%%%%%%%%%%%%%%%%%%%%%%%%%%%%%%%%%%%%%%%%%%%%%%
For $a$ near zero  we neglect the term proportional to $a^4$ in  $\mathrm{U}(k,a)$,
\begin{equation}
	\label{u-}
	\mathrm{U}(k,a)\sim U_-(k,a)=\frac{a^2}{\lambda^2}\left(1-\frac{a_-(k)^4}{a^4} \right),
	\quad
	\mbox{ as } a \to 0^+,
\end{equation}
where
\begin{equation}\label{a-}
a_-(k)=(\sqrt{6}{M_{\rm pl}} \lambda |k|)^{1/2},
\end{equation}
is the unique positive zero of  $U_-(k,a)$. The superpotential $U_-(k,a)$, illustrated in Fig.~4,
corresponds to quantum tunneling  between the classically allowed region $0<a<a_-(k)$
and the classically forbidden region $a_-(k)<a$, The analog of Eq.~(\ref{WDW1+}) is
\begin{equation}
	\label{WDW1-}
	\left(  \frac{d^2}{d a^2}+\frac{p}{a}\frac{d}{d a}-\frac{1}{\hbar^2}U_-(k,a) \right) \psi^{(0)}(k,a) = 0.
\end{equation}

%%%%%%%%%%%%%%%%%%%%%%%%%%%%%%%%%%%%%%%%%%%%%%%%%%%%%%%%%%%%%%%%%
\begin{figure}
\begin{center}
        \includegraphics[width=8cm]{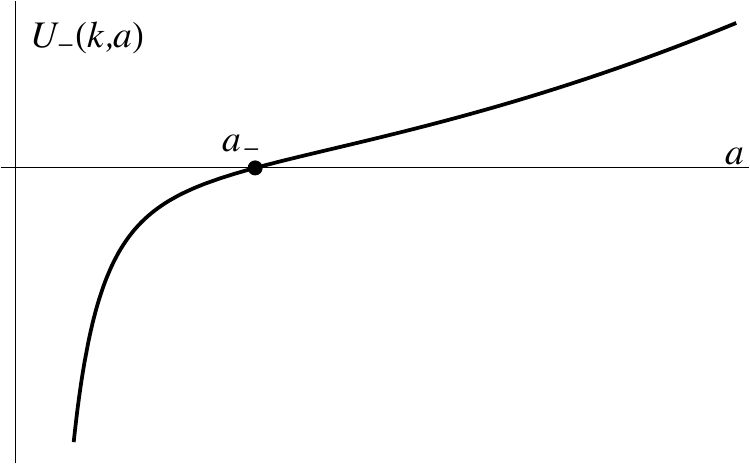}
 \end{center}
    \caption{Generic shape of the superpotential $U_-(k,a)$ (with $k\neq 0$) defined in Eqs.~(\ref{u-})--(\ref{a-}).}
\end{figure}
%%%%%%%%%%%%%%%%%%%%%%%%%%%%%%%%%%%%%%%%%%%%%%%%%%%%%%%%%%%%%%%%%

Again, since $a_-(k)$ is the unique positive zero of $U_-(k,a)$, it is natural to take it as the reference
point for the WKB solutions. By using Eq.~(\ref{ca2})  with $\mathrm{U}(k,a)$ replaced by $U_-(k,a)$
and $a_r=a_-(k)$, we obtain the following approximations $\psi_{\pm}^{(0)}(k,a)$
to the functions $\psi_{\pm}(k,a)$ valid for small $a$,
\begin{equation}
	\label{wke-}
	\psi_{\pm}^{(0)}(k,a) =
	\lambda^{1/2}a^{-\frac{p+1}{2}}\left(\frac{a_-(k)^4}{a^4} -1\right)^{-1/4}
	\exp\left( \pm \frac{i}{\hbar}\int_{a_-(k)}^a \mathrm{p}_-(k,a') d a' \right),
\end{equation}
where
\begin{eqnarray}
	\int_{a_-(k)}^a \mathrm{p}_-(k,a') d a'
	& = & \frac{1}{2\lambda}
	          \left[\sqrt{a_-(k)^4-a^4}+a_-(k)^2 \log a^2 \right.\nonumber\\
	\label{comp}	   
	&  & \qquad{}\left.-a_-(k)^2\log\left( a_-(k)^2+\sqrt{a_-(k)^4-a^4}\right)\right].
\end{eqnarray}
Since for $a\ll a_-(k)$
\begin{equation}
	\label{che}
	\int_{a_-(k)}^a \mathrm{p}_-(k,a') d a'\sim \frac{a_-(k)^2}{\lambda} \log a,
\end{equation}
Eq.~(\ref{wke-}) implies
\begin{equation}
	\label{psit}
	\psi_{\pm}^{(0)}(k,a)
	\sim
	\frac{\lambda^{1/2}}{a_-(k)} a^{(1-p)/2} \exp\left( \pm\frac{i}{\hbar}\sqrt{6}\, M_{\rm pl} |k|\log a \right),\quad a\ll a_-(k).
\end{equation}
Thus the complete wavefunctions can be approximated as
\begin{equation}
	\label{one}
	\Psi_{\pm}^{(0)}(k,a,\phi)
	\sim
	\frac{\lambda^{1/2}}{a_-(k)} a^{2\nu}
	\exp\left[ \pm\frac{i}{\hbar}(\sqrt{6}\, M_{\rm pl} |k|\log a \pm k\phi) \right], \quad a\ll a_-(k),
\end{equation}
where  $\nu=(1-p)/4$. The admissible wavefunctions should  be  regular  at $a=0$ and this is satisfied
only if $\nu\geq 0$ or, equivalently, if $p\leq 1$.

The  wavefronts of these approximations are given by the classical trajectories of $(a,\phi)$ in the  KD period,
\begin{equation}
	\label{wfr}
	\sqrt{6}\,M_{\rm pl} \log a \pm \phi=const,
\end{equation}
and the wavefunctions $\Psi_{+}^{(0)}(k,a,\phi)$ and $\Psi_{-}^{(0)}(k,a,\phi)$ represent  contracting  and expanding universes
in the KD period, respectively.

Finally, taking into account Eq.~(\ref{pa2r}), from Eq.~(\ref{a-}) we find that
\begin{equation}
	\label{des-}
 	a_-^2 = \frac{ |k|}{2\sqrt{6}\pi^2 M_{\rm pl}}.
\end{equation}
Therefore, $a_-$ satisfies the classical limit condition Eq.~(\ref{bo}) for the scale factor if
\begin{equation}
	\label{clasc}
	|k| > \frac{6\sqrt{6}\pi^2}{M_{\rm pl}}.
\end{equation}
%%%%%%%%%%%%%%%%%%%%%%%%%%%%%%%%%%%%%%%%%%%%%%%%%%%%%%%%%%%%%%%%%
\section{TYPES OF EMERGENT UNIVERSES}
%%%%%%%%%%%%%%%%%%%%%%%%%%%%%%%%%%%%%%%%%%%%%%%%%%%%%%%%%%%%%%%%%
In this Section we will describe the different types of classical universes which may emerge from the solutions
Eq.~(\ref{nac}) of the WDW equation with a constant potential.
%%%%%%%%%%%%%%%%%%%%%%%%%%%%%%%%%%%%%%%%%%%%%%%%%%%%%%%%%%%%%%%%%
\subsection{Type-$A$ universes}
%%%%%%%%%%%%%%%%%%%%%%%%%%%%%%%%%%%%%%%%%%%%%%%%%%%%%%%%%%%%%%%%%
Classical universes evolving according to solutions of type $A_1$ (expanding)  and $A_2$ (contracting) of
Eqs.~(\ref{co1})--(\ref{co2}) with both KD and INF periods arise provided that $(4/27) a_+^4 < b_+$,
which restated in terms of the parameters appearing in the corresponding semiclassical wavefunctions
is equivalent to (see Eq.~(\ref{co3b})),
\begin{equation}
	\label{typea}
	k^2 > 32\pi^4\frac{{M_{\rm pl}}^6}{V_0^2}.
\end{equation}
In this case $\mathrm{U}(k,a)$ is strictly negative on the domain $a>0$, and the wavefunctions
\begin{equation}
	\label{psA}
	\Psi_{\pm}(k,a,\phi) = \psi_{\pm}(k,a) e^{ i k\phi/\hbar},
\end{equation}
where
\begin{equation}
	\label{ca2b}
	\psi_{\pm}(k,a)= \frac{1}{(a^p\mathrm{p}(k,a))^{1/2}} \exp \left( \pm \frac{i}{\hbar}\int_{a_+}^a \mathrm{p}(k,a') d a' \right),
\end{equation}
are well-defined oscillatory functions on the whole  classically allowed domain $a>0$.

In Sec.~IV-B we showed that for large $a$ the function $\Psi_{-}(k,a,\phi)$ describes expanding INF universes.
If we factorize $\psi_-(k,a)$ in the form
\begin{equation}
	\label{dec}
	\psi_{-}(k,a)
	=
	\frac{1}{(a^p\mathrm{p}(k,a))^{1/2}}
	\exp \left( - \frac{i}{\hbar}\int_{a_+}^{a_-(k)} \mathrm{p}(k,a') d a' \right)
	\exp \left( - \frac{i}{\hbar}\int_{a_-(k)}^a \mathrm{p}(k,a') d a' \right),
\end{equation}
we see that near $a=0$ the last factor behaves as $\psi_-^{(0)}(k,a)$ and, from the analysis of
Sec.~IV-C,  in this small $a$ region $\Psi_{-}(k,a,\phi)$ describes expanding universes in the KD regime.
Therefore $\Psi_{-}(k,a,\phi)$ is indeed peaked about a type-$A_1$ solution.

A similar argument shows that  $\Psi_{+}(k,a,\phi)$ is peaked about a type-$A_2$ solution.
%%%%%%%%%%%%%%%%%%%%%%%%%%%%%%%%%%%%%%%%%%%%%%%%%%%%%%%%%%%%%%%%%
\subsection{Type-$B$ universes}
%%%%%%%%%%%%%%%%%%%%%%%%%%%%%%%%%%%%%%%%%%%%%%%%%%%%%%%%%%%%%%%%%
Classical universes evolving according to solutions of type $B_1$ or $B_2$ of  Eqs.~(\ref{co1})--(\ref{co2}) arise if
$0 < b_+ < (4/27) a_+^4$, or, equivalently,
\begin{equation}
	\label{cc}
	0<k^2< 32\pi^4\frac{{M_{\rm pl}}^6}{V_0^2}.
\end{equation}
In this case $\mathrm{U}(k,a)$ has two positive zeros $0<a_1<a_2$
given by Eq.~(\ref{solap0}), and corresponds to the potential barrier illustrated in Fig.~5.
%%%%%%%%%%%%%%%%%%%%%%%%%%%%%%%%%%%%%%%%%%%%%%%%%%%%%%%%%%%%%%%%%
\begin{figure}
\begin{center}
        \includegraphics[width=8cm]{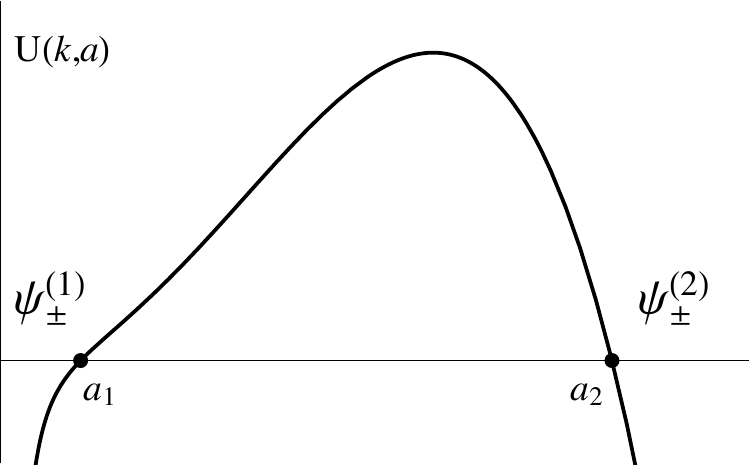}
 \end{center}
    \caption{Generic shape of the superpotential $\mathrm{U}(k,a)$ for $0<k^2< 32\pi^4 {M_{\rm pl}}^6/V_0^2$ with
                  the two classically allowed regions $0<a<a_1$ and $a_2<a$.}
\end{figure}
%%%%%%%%%%%%%%%%%%%%%%%%%%%%%%%%%%%%%%%%%%%%%%%%%%%%%%%%%%%%%%%%%

There are two classically allowed regions, $0<a<a_1$ and $a_2<a$, with respective
 basic WKB wavefunctions given by,
 \begin{equation}
 	\label{wB2}
	\psi_{\pm}^{(r)}(k,a)
	=
	\frac{1}{(a^p\mathrm{p}(k,a))^{1/2}}
	\exp \left( \pm \frac{i}{\hbar}\int_{a_r}^a \mathrm{p}(k,a') d a' \right),\quad r=1, 2,
\end{equation}
which in turn can be approximated by $\psi_{\pm}^{(0)}(k,a)$ and $\psi_{\pm}^{(\infty)}(a)$ for $r=1$ and $r=2$ respectively.

Using the standard WKB connection formulas (wherein the semiclassical expansions are interpreted as asymptotic
expansions in the sense of Poincar\'e) to connect these two sets of WKB solutions across the barrier
leads to a well-known loss of unitarity~\cite{BM72,GA91}. To circumvent this problem, we interpret the asymptotic
expansions in the sense of Borel, and use the ensuing connections formulas of Silverstone~\cite{SI85}, which
follow from writing the exact wavefunction in Langer-Cherry form~\cite{LA37,CH50},
\begin{equation}
	\psi(a) = 2\pi^{1/2} \hbar^{-1/6} (\dot{\phi}(a,\hbar))^{-1/2}
              [\alpha \mathrm{Ai} (-\hbar^{2/3}\phi(a,\hbar)) +  \beta \mathrm{Bi}(-\hbar^{2/3}\phi(a,\hbar))],
\end{equation}
expanding the new independent variable $\phi(a,\hbar)$ as a power series in $\hbar^2$, and using the Borel
summability of the Airy functions $\mathrm{Ai}(z)$ and $\mathrm{Bi}(z)$.
Since the classically forbidden region is a Stokes line, it is necessary to ``pick sides,'' and we choose to take
$\mathrm{Im}\,a\to 0^+$. The calculation is analogous to the calculation corresponding to the piecewise
linear potential in Fig.~1(b) of Ref.~\cite{SS04}, where all the three intermediate steps are carried out in detail:
(i) connecting the properly normalized semiclassical wavefunctions across $a_1$; (ii) changing the reference
point of the semiclassical wavefunctions to $a_2$; and (iii) connecting the properly normalized semiclassical
wavefunctions across $a_2$. The connection formulas themselves are given in Eqs.~(37)--(40)
of Ref.~\cite{SS04}. Using them we find,
\begin{equation}
	\psi_{+}^{(2)}(k,a) \rightarrow e^{ K(k)} (\psi_{+}^{(1)}(k,a)- i \psi_{-}^{(1)}(k,a) ),
\end{equation}
\begin{equation}
	\psi_{-}^{(2)}(k,a) \rightarrow e^{ K(k)} (\psi_{-}^{(1)}(k,a) + i\,\psi_{+}^{(1)}(k,a) )
	                                              - i e^{-K(k)} \psi_{+}^{(1)}(k,a),
\end{equation}
and, conversely,
\begin{equation}
	\psi_{+}^{(1)}(k,a)\rightarrow e^{ K(k)} ( \psi_{+}^{(2)}(k,a)+ i \psi_{-}^{(2)} (k,a) ),
\end{equation}
\begin{equation}
	\psi_{-}^{(1)}(k,a)\rightarrow e^{ K(k)} (\psi_{-}^{(2)}(k,a) - i \psi_{+}^{(2)} (k,a) )
	                                             + i e^{-K(k)} \psi_{+}^{(2)}(k,a),
\end{equation}
where
\begin{equation}
	\label{fac}
	K(k)=\frac{1}{\hbar}\int_{a_1}^{a_2}| \mathrm{p}(k,a)|\, da.
\end{equation}

As  a consequence it follows that
\begin{equation}
	\label{01}
	\psi_{+}^{(1)}(k,a)-i\psi_{-}^{(1)}(k,a) \rightarrow e^{-K(k)} \psi_{+}^{(2)}(k,a),
\end{equation}
and
\begin{equation}
	\label{02}
	\psi_{+}^{(2)}(k,a)+i\psi_{-}^{(2)}(k,a) \rightarrow e^{-K(k)} \psi_{+}^{(1)}(k,a).
\end{equation}
(The standard WKB connection formulas would have led to Eqs.~(\ref{01}) and~(\ref{02}) with
their right-hand sides equal to zero, with the apparent result that a nonzero oscillatory wavefunction
at one side of the barrier connects to a zero wavefunction on the other side of the barrier. These
connection formulas do not capture the physically intuitive result that the wavefunction on the other
side of the barrier is also oscillatory but suppressed by an exponentially small factor of the order
of $\exp(-K(k))$.)

Therefore, the wavefunction corresponding to the WKB approximation given by
\begin{equation}
	\label{01a}
	\Psi^{(1)}(k,a,\phi) = \left( \psi_{+}^{(1)}(k,a)-i\psi_{-}^{(1)}(k,a)\right) e^{i k\phi/\hbar},\quad  0<a< a_1,
\end{equation}
is a superposition of an expanding component and a contracting component peaked about classical solutions of
Eqs.~(\ref{co1})--(\ref{co2}) in a KD period. In addition, this WKB wavefunction is suppressed by an exponentially small
factor for $a>a_2$ and, consequently, $\Psi^{(1)}(k,a,\phi)$ is peaked about $B_1$ solutions of Eqs.~(\ref{co1})--(\ref{co2}).

Likewise, the wavefunction corresponding to the WKB approximation
\begin{equation}
	\label{01a2}
	\Psi^{(2)}(k,a,\phi)=\left( \psi_{+}^{(2)}(k,a)+i\psi_{-}^{(2)}(k,a)\right) e^{i k\phi/\hbar},\quad  a_2<a< \infty,
\end{equation}
is a superposition of expanding and contracting components peaked about classical solutions of Eqs.~(\ref{co1})--(\ref{co2})
in an inflationary period, and is suppressed by an exponentially small factor for $ 0<a<a_1$ .
Therefore $\Psi^{(2)}(k,a,\phi)$ is peaked about the $B_2$ type solutions of Eqs.~(\ref{co1})--(\ref{co2}).
Note that for large $a$ the no-boundary wavefunction $\psi_\mathrm{NB}(k,a,\phi)$ and $\Psi^{(2)}(k,a,\phi)$ are related by
\begin{equation}
	\label{prop}
	\Psi_\mathrm{NB}(k,a,\phi) = A^{-1} e^{-i\pi/4}\Psi^{(2)}(k,a,\phi).
\end{equation}
%%%%%%%%%%%%%%%%%%%%%%%%%%%%%%%%%%%%%%%%%%%%%%%%%%%%%%%%%%%%%%%%%
\subsection{Type-$C$ and type-$D$ universes}
%%%%%%%%%%%%%%%%%%%%%%%%%%%%%%%%%%%%%%%%%%%%%%%%%%%%%%%%%%%%%%%%%
Type-$C$  universes emerge as semiclassical limits of wavefunctions for  $k^2=32\pi^4 {M_{\rm pl}}^6/V_0^2$, and are
limiting cases of type-$B$ universes as $a_1\rightarrow a_2$.

Finally, type-$D$ universes are associated with wavefunctions with $k=0$, and are the same as those  studied in Ref.~\cite{VI88}.
The subtype $D_1$ describes expanding inflationary universes without a KD period and corresponds to
$\psi_{-}^{(\infty)}(a)$ wavefunctions in Eq.~(\ref{wke}). Subtype $D_2$  describes contracting universes without
a KD period and corresponds to wavefunctions $\psi_{+}^{(\infty)}(a)$ in Eq.~(\ref{wke}).
%%%%%%%%%%%%%%%%%%%%%%%%%%%%%%%%%%%%%%%%%%%%%%%%%%%%%%%%%%%%%%%%%
\section{PROBABILITY DISTRIBUTIONS}
%%%%%%%%%%%%%%%%%%%%%%%%%%%%%%%%%%%%%%%%%%%%%%%%%%%%%%%%%%%%%%%%%
The probabilistic interpretation of the solutions of the WDW equation is formulated in terms of the
components of the current~\cite{DE67,VI86,VI88,ha09}
\begin{equation}
	\label{cura}
	j^a = \frac{i}{2} \hbar \, a^p( \Psi^*\partial_a \Psi-\Psi \partial_a \Psi^*),
\end{equation}
\begin{equation}
	\label{curf}
	 j^{\phi} =- i \, 3{M_{\rm pl}}^2 \, \hbar\,a^{p-2}( \Psi^*\partial_\phi \Psi-\Psi \partial_\phi \Psi^*),
\end{equation}
which satisfy the continuity equation
\begin{equation}
	\label{cur2}
	\partial_a j^a + \partial_\phi j^{\phi}=0.
\end{equation}
In general, these components are not positive definite and their interpretation requires
the analysis of an appropriate time variable~\cite{VI86,VI88}.
If $a$ is taken as a time variable, then $\rho_{\phi_0}(a)= j^a$ is interpreted as the probability
density for $\phi=\phi_0$ at a given value of $a$. The quantity $\partial_{\phi}j^{\phi}$ for the solutions
$\Psi(k,a,\phi)$ of the form of Eq.~(\ref{nac}) vanishes identically and, consequently, the density $\rho_{\phi_0}(a)$
is $a$-independent. It turns out that the $\rho_{\phi_0}(a)$ for a generic WKB wavefunction of the WDW equation
is positive (negative) if the wavefunction corresponds to expanding (contracting) universes~\cite{VI86}.

Let us consider the expanding universes with an inflationary period $A_1, B_2,C_3$ and $D_1$.
Their wavefunctions for large $a$ are approximated by $\psi_{-}^{(\infty)}(a)$, and from Eq.~(\ref{ca2}) it follows that
\begin{equation}
	\label{sisi0}
	\psi_{-}^{(\infty)*}\partial_{a} \psi_{-}^{(\infty)}-\psi_{-}^{(\infty) } \partial_{a} \psi_{-}^{(\infty) *}=- \frac{2i} {\hbar a^p},
\end{equation}
which leads to a constant density
\begin{equation}
	\label{re}
	\rho_{\phi_0}(a)= 1.
\end{equation}

As a consequence of Eq.~(\ref{re}), Vilenkin's tunneling wavefunction $\psi_\mathrm{T}$ on the  region $a>a_+$ satisfies
\begin{equation}
	\label{amp}
	\rho_{T,\phi_0}(a) = \exp{\left(-\frac{24 \pi^2 {M_{\rm pl}}^4 }{\hbar V(\phi_0)}\right)}=\exp{\left(-\frac{2}{3\mathbb{V}(\phi_0)}\right)},
\end{equation}
where $\mathbb{V}=\hbar V/(6\pi {M_{\rm pl}}^2)^2$ is the potential function with the Vilenkin normalization
(see Eq.~(4.15) in Ref.~\cite{VI88}).  This expression is  the standard result for the probability density
for the Vilenkin tunneling function~\cite{VI88,ha09}.

Of the four types of universes $A_1, B_2,C_3$ and $D_1$ under consideration, only the type $A_1$ has a KD period
in addition to the inflationary period. Moreover, from Eq.~(\ref{dec}) we have that near $a=0$ the wavefunction corresponding
to the $A_1$ type is approximated by
\begin{equation}
	\label{decb}
	\psi_{-}(k,a)=  \exp \left( - \frac{i}{\hbar}\int_{a_+}^{a_-(k)} \mathrm{p}(k,a') d a' \right) \psi_-^{(0)}(k,a).
\end{equation}
Then, using Eqs.~(\ref{one})  and~(\ref{a-})  we obtain
\begin{equation}
	\label{dens1}
	\rho_{\phi_0}(a) = \frac{\lambda}{a_-^2} \sqrt{6}{M_{\rm pl}} |k| =1
\end{equation}
which is the same constant density as that calculated for large $a$ in Eq.~(\ref{re}).
%%%%%%%%%%%%%%%%%%%%%%%%%%%%%%%%%%%%%%%%%%%%%%%%%%%%%%%%%%%%%%%%%
\section{CONCLUSIONS}
%%%%%%%%%%%%%%%%%%%%%%%%%%%%%%%%%%%%%%%%%%%%%%%%%%%%%%%%%%%%%%%%%
In the present work we have discussed the emergence of KD classical universes in closed Friedmann universes with
unit curvature from a certain family of WKB solutions of the WDW equation with a constant inflaton potential.
These WKB solutions are semiclassical approximations to eigenfunctions of the momentum operator conjugate to the inflaton,
i.e., they depend on the inflaton field $\phi$ through a phase factor $\exp{i(k\phi/\hbar)}$. The standard treatment of this
problem has been restricted to $k=0$, and leads to a large-$a$, classically allowed, INF region, and to a small-$a$,
classically forbidden region. Allowing for $k\neq 0$ replaces this scenario with a new one, in which two classically allowed
regions separated by a potential barrier via two turning points may arise. In the innermost region there are semiclassical
solutions peaked about KD classical solutions, and in the outermost region we find, among others, the well-known
Vilenkin tunneling and Hartle-Hawking no-boundary solutions times the $\phi$-dependent phase factor. 
We have also discussed how these solutions connect in our setting to superpositions of semiclassical
wavefunctions in the KD regime. We have recovered within our approach the standard results on the INF region.
Finally,  in the Appendix we use the explicit integrability of the WDW equation near $a=0$ to provide expressions for the wavefunctions
with WKB approximations  on the KD regions in terms of first-kind modified Bessel functions.

Future work will involve alternative models for the possible emergence of classical universes
which arise from the quantization of  Friedmann equations exhibiting clasically allowed regions for small values
of the scale factor. Appropriate candidates are, e.g., the inflaton models originated  from cosmology  driven by conformal field theory
presented in Refs.~\cite{BAR15,BAR16}, wherein the effective potentials in the corresponding  Friedmann equations
indeed turn out to be negative at small scale factors.
%%%%%%%%%%%%%%%%%%%%%%%%%%%%%%%%%%%%%%%%%%%%%%%%%%%%%%%%%%%%%%%%%
\begin{acknowledgments}
The financial support of the Spanish Ministerio de Econom\'{\i}a y Competitividad
under Project No.~PGC2018-094898-B-I00 is gratefully acknowledged.
\end{acknowledgments}
%%%%%%%%%%%%%%%%%%%%%%%%%%%%%%%%%%%%%%%%%%%%%%%%%%%%%%%%%%%%%%%%%
\appendix
%%%%%%%%%%%%%%%%%%%%%%%%%%%%%%%%%%%%%%%%%%%%%%%%%%%%%%%%%%%%%%%%%
\section{Solution of the WDW equation near $a=0$ in terms of Bessel functions}
%%%%%%%%%%%%%%%%%%%%%%%%%%%%%%%%%%%%%%%%%%%%%%%%%%%%%%%%%%%%%%%%%
The general solution of Eq.~(\ref{WDW1-}) is a linear combination of the functions
\begin{equation}
	\label{bess}
	f_{\pm}(k,a)=I_{\pm \omega(k)}(a^2/2\hbar\lambda)a^{2\nu},
\end{equation}
where $I_{\pm\omega}$ are the first-kind modified Bessel functions ~\cite{AS67} and
\begin{equation}
	\label{alp}
 	\omega(k) = \frac{1}{2} \left[(2\nu)^2-6{M_{\rm pl}}^2  k^2/\hbar^2\right]^{1/2}.
\end{equation}
The generic shape of the real and imaginary parts of Eq.~(\ref{bess}) are illustrated in Fig.~6.
From the asymptotic formula~\cite{AS67}
\begin{equation}
	\label{ass2}
	I_{\omega}(x) \sim \frac{1}{\Gamma(1+\omega)}\left(\frac{x}{ 2}\right)^{\omega},
	\quad\mbox{ as } x\rightarrow 0,
\end{equation}
it follows that
\begin{equation}
	\label{ass3}
	f_{\pm}(k,a) \sim C_{\pm} a^{2(\nu\pm \omega(k))},
	\quad\mbox{ as } a\rightarrow 0,
\end{equation}
where
 \begin{equation}
 	\label{wkbn2}
	C_{\pm}= \frac{1}{\Gamma(1\pm \omega(k)) (4  \hbar\lambda)^{\pm \omega(k)}}.
\end{equation}

A necessary  condition for the validity of the WKB approximation Eq.~(\ref{wke-}) is
 \begin{equation}
 	\label{unn2}
       \frac{6M_{\rm pl}^2 k^2}{\hbar^2} \gg 4\nu^2,
 \end{equation}
 in which case $\omega(k)$ is purely imaginary and in fact $\omega(k)\sim i\sqrt{6}M_{\rm pl} |k|/(2\hbar)$.
Therefore, using Eq.~(\ref{ass3}) for functions $f_{\pm}(k,a)$ with values of $k$ satisfying Eq.~(\ref{unn2}),
it follows that
\begin{eqnarray}
	F_{\pm}(k,a,\phi)
	& = &
	f_{\pm} (k,a) e^{ i k \phi/\hbar } \\
 	\label{wkbn1}
	& \sim &
	C_{\pm} a^{2\nu}\exp \left[ \pm \frac{i}{\hbar}\,(\sqrt{6}\,M_{\rm pl} |k|\log a \pm k\phi)\right],
	\quad\mbox{ as } a\rightarrow 0.
\end{eqnarray}
Therefore, in view of Eqs.~(\ref{one}) and~(\ref{wkbn1}) it is clear that as $ a\rightarrow 0$ the 
wavefunctions $\Psi_{\mp}^{(0)}(k,a,\phi)$ are proportional to  the semiclassical expansions of the explicit functions
\begin{equation}
	\label{id}
	F_{\pm}(k,a,\phi) = I_{\pm \omega(k)}(a^2/2\hbar\lambda)a^{2\nu}e^{i k \phi/\hbar}.
\end{equation}

%%%%%%%%%%%%%%%%%%%%%%%%%%%%%%%%%%%%%%%%%%%%%%%%%%%%%%%%%%%%%%%%%
\begin{figure}
\begin{center}
        \includegraphics[width=7cm]{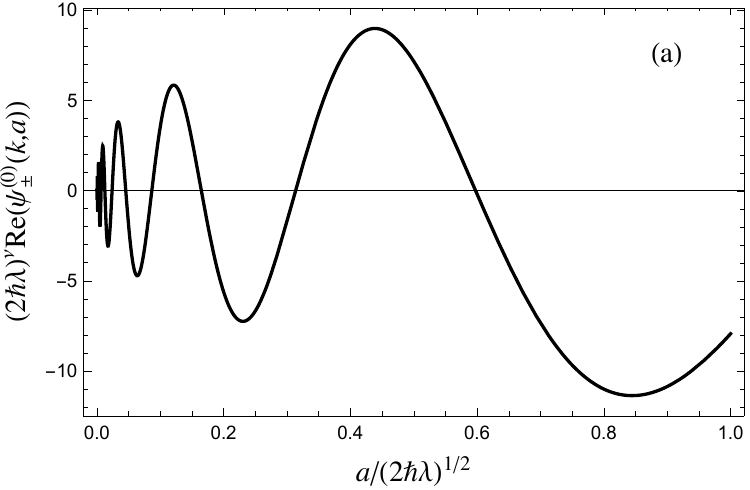}
        \hspace{1cm}
        \includegraphics[width=7cm]{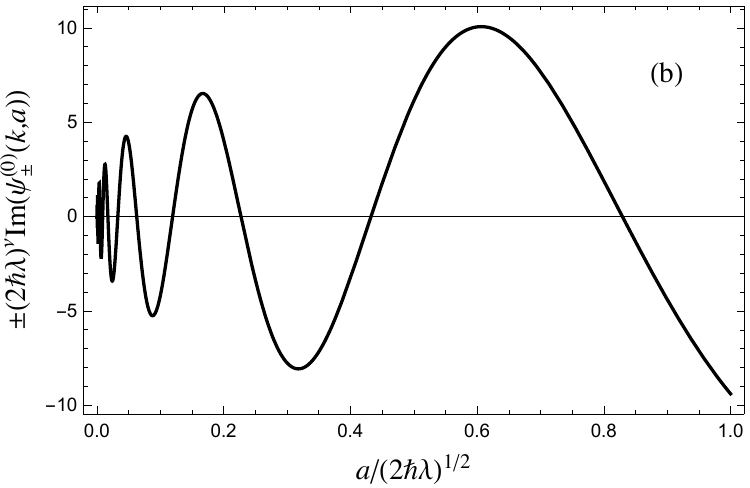}
\end{center}
    \caption{The functions (a) $(2\hbar \lambda)^{\nu} \mathrm{Re}(\psi_{\pm}^{(0)}(k,a))$,
                  and (b)  $\pm (2\hbar \lambda)^{\nu} \mathrm{Im}(\psi_{\pm}^{(0)}(k,a))$
                  on the oscillatory region near $a=0$ for $\nu=1/6$ and $k=2 \hbar/{M_{\rm pl}}$.}
\end{figure}
%%%%%%%%%%%%%%%%%%%%%%%%%%%%%%%%%%%%%%%%%%%%%%%%%%%%%%%%%%%%%%%%%
%%%%%%%%%%%%%%%%%%%%%%%%%%%%%%%%%%%%%%%%%%%%%%%%%%%%%%%%%%%%%%%%%
\bibliography{kd}

\begin{thebibliography}{41}%
\makeatletter
\providecommand \@ifxundefined [1]{%
 \@ifx{#1\undefined}
}%
\providecommand \@ifnum [1]{%
 \ifnum #1\expandafter \@firstoftwo
 \else \expandafter \@secondoftwo
 \fi
}%
\providecommand \@ifx [1]{%
 \ifx #1\expandafter \@firstoftwo
 \else \expandafter \@secondoftwo
 \fi
}%
\providecommand \natexlab [1]{#1}%
\providecommand \enquote  [1]{``#1''}%
\providecommand \bibnamefont  [1]{#1}%
\providecommand \bibfnamefont [1]{#1}%
\providecommand \citenamefont [1]{#1}%
\providecommand \href@noop [0]{\@secondoftwo}%
\providecommand \href [0]{\begingroup \@sanitize@url \@href}%
\providecommand \@href[1]{\@@startlink{#1}\@@href}%
\providecommand \@@href[1]{\endgroup#1\@@endlink}%
\providecommand \@sanitize@url [0]{\catcode `\\12\catcode `\$12\catcode
  `\&12\catcode `\#12\catcode `\^12\catcode `\_12\catcode `\%12\relax}%
\providecommand \@@startlink[1]{}%
\providecommand \@@endlink[0]{}%
\providecommand \url  [0]{\begingroup\@sanitize@url \@url }%
\providecommand \@url [1]{\endgroup\@href {#1}{\urlprefix }}%
\providecommand \urlprefix  [0]{URL }%
\providecommand \Eprint [0]{\href }%
\providecommand \doibase [0]{https://doi.org/}%
\providecommand \selectlanguage [0]{\@gobble}%
\providecommand \bibinfo  [0]{\@secondoftwo}%
\providecommand \bibfield  [0]{\@secondoftwo}%
\providecommand \translation [1]{[#1]}%
\providecommand \BibitemOpen [0]{}%
\providecommand \bibitemStop [0]{}%
\providecommand \bibitemNoStop [0]{.\EOS\space}%
\providecommand \EOS [0]{\spacefactor3000\relax}%
\providecommand \BibitemShut  [1]{\csname bibitem#1\endcsname}%
\let\auto@bib@innerbib\@empty
%</preamble>
\bibitem [{\citenamefont {Starobinsky}(1980)}]{STA80}%
  \BibitemOpen
  \bibfield  {author} {\bibinfo {author} {\bibfnamefont {A.}~\bibnamefont
  {Starobinsky}},\ }\bibfield  {title} {\bibinfo {title} {A new type of
  isotropic cosmological models without singularity},\ }\href@noop {}
  {\bibfield  {journal} {\bibinfo  {journal} {Phys. Lett. B}\ }\textbf
  {\bibinfo {volume} {91}},\ \bibinfo {pages} {99} (\bibinfo {year}
  {1980})}\BibitemShut {NoStop}%
\bibitem [{\citenamefont {Guth}(1981)}]{GU81}%
  \BibitemOpen
  \bibfield  {author} {\bibinfo {author} {\bibfnamefont {A.~H.}\ \bibnamefont
  {Guth}},\ }\bibfield  {title} {\bibinfo {title} {Inflationary universe: A
  possible solution to the horizon and flatness problems},\ }\href@noop {}
  {\bibfield  {journal} {\bibinfo  {journal} {Phys. Rev. D}\ }\textbf {\bibinfo
  {volume} {23}},\ \bibinfo {pages} {347} (\bibinfo {year} {1981})}\BibitemShut
  {NoStop}%
\bibitem [{\citenamefont {Linde}(1982)}]{LI82}%
  \BibitemOpen
  \bibfield  {author} {\bibinfo {author} {\bibfnamefont {A.~D.}\ \bibnamefont
  {Linde}},\ }\bibfield  {title} {\bibinfo {title} {A new inflationary universe
  scenario: A possible solution of the horizon, flatness, homogeneity, isotropy
  and primordial monopole problems},\ }\href@noop {} {\bibfield  {journal}
  {\bibinfo  {journal} {Phys. Lett. B}\ }\textbf {\bibinfo {volume} {108}},\
  \bibinfo {pages} {389} (\bibinfo {year} {1982})}\BibitemShut {NoStop}%
\bibitem [{\citenamefont {Linde}(1985)}]{LI85}%
  \BibitemOpen
  \bibfield  {author} {\bibinfo {author} {\bibfnamefont {A.~D.}\ \bibnamefont
  {Linde}},\ }\bibfield  {title} {\bibinfo {title} {Initial conditions for
  inflation},\ }\href@noop {} {\bibfield  {journal} {\bibinfo  {journal} {Phys.
  Lett. B}\ }\textbf {\bibinfo {volume} {162}},\ \bibinfo {pages} {281}
  (\bibinfo {year} {1985})}\BibitemShut {NoStop}%
\bibitem [{\citenamefont {Mukhanov}(2005)}]{MU05}%
  \BibitemOpen
  \bibfield  {author} {\bibinfo {author} {\bibfnamefont {V.}~\bibnamefont
  {Mukhanov}},\ }\href@noop {} {\emph {\bibinfo {title} {Physical Foundations
  of Cosmology}}}\ (\bibinfo  {publisher} {Cambridge University Press},\
  \bibinfo {year} {2005})\BibitemShut {NoStop}%
\bibitem [{\citenamefont {Baumann}(2009)}]{BA09}%
  \BibitemOpen
  \bibfield  {author} {\bibinfo {author} {\bibfnamefont {D.}~\bibnamefont
  {Baumann}},\ }\bibfield  {title} {\bibinfo {title} {Tasi {L}ectures on
  {I}nflation},\ }\href@noop {} {\bibfield  {journal} {\bibinfo  {journal}
  {arXiv:0907.5424}\ } (\bibinfo {year} {2009})}\BibitemShut {NoStop}%
\bibitem [{\citenamefont {Martin}(2018)}]{MA18}%
  \BibitemOpen
  \bibfield  {author} {\bibinfo {author} {\bibfnamefont {J.}~\bibnamefont
  {Martin}},\ }\bibfield  {title} {\bibinfo {title} {{The Theory of
  Inflation}},\ }in\ \href@noop {} {\emph {\bibinfo {booktitle} {{200th Course
  of Enrico Fermi School of Physics}: {Gravitational Waves and Cosmology}}}}\
  (\bibinfo {year} {2018})\ \Eprint {https://arxiv.org/abs/1807.11075}
  {arXiv:1807.11075 [astro-ph.CO]} \BibitemShut {NoStop}%
\bibitem [{\citenamefont {Hartle}\ and\ \citenamefont {Hawking}(1983)}]{HH83}%
  \BibitemOpen
  \bibfield  {author} {\bibinfo {author} {\bibfnamefont {J.~B.}\ \bibnamefont
  {Hartle}}\ and\ \bibinfo {author} {\bibfnamefont {S.~W.}\ \bibnamefont
  {Hawking}},\ }\bibfield  {title} {\bibinfo {title} {Wave function of the
  universe},\ }\href@noop {} {\bibfield  {journal} {\bibinfo  {journal} {Phys.
  Rev. D}\ }\textbf {\bibinfo {volume} {28}},\ \bibinfo {pages} {2960}
  (\bibinfo {year} {1983})}\BibitemShut {NoStop}%
\bibitem [{\citenamefont {Hawking}(1984)}]{H84}%
  \BibitemOpen
  \bibfield  {author} {\bibinfo {author} {\bibfnamefont {S.~W.}\ \bibnamefont
  {Hawking}},\ }\bibfield  {title} {\bibinfo {title} {The quantum state of the
  universe},\ }\href@noop {} {\bibfield  {journal} {\bibinfo  {journal}
  {Nuclear Physics B}\ }\textbf {\bibinfo {volume} {239}},\ \bibinfo {pages}
  {257} (\bibinfo {year} {1984})}\BibitemShut {NoStop}%
\bibitem [{\citenamefont {Vilenkin}(1986)}]{VI86}%
  \BibitemOpen
  \bibfield  {author} {\bibinfo {author} {\bibfnamefont {A.}~\bibnamefont
  {Vilenkin}},\ }\bibfield  {title} {\bibinfo {title} {Boundary conditions in
  quantum cosmology},\ }\href@noop {} {\bibfield  {journal} {\bibinfo
  {journal} {Phys. Rev. D}\ }\textbf {\bibinfo {volume} {33}},\ \bibinfo
  {pages} {3560} (\bibinfo {year} {1986})}\BibitemShut {NoStop}%
\bibitem [{\citenamefont {Vilenkin}(1988)}]{VI88}%
  \BibitemOpen
  \bibfield  {author} {\bibinfo {author} {\bibfnamefont {A.}~\bibnamefont
  {Vilenkin}},\ }\bibfield  {title} {\bibinfo {title} {Quantum cosmology and
  the initial state of the universe},\ }\href@noop {} {\bibfield  {journal}
  {\bibinfo  {journal} {Phys. Rev. D}\ }\textbf {\bibinfo {volume} {37}},\
  \bibinfo {pages} {888} (\bibinfo {year} {1988})}\BibitemShut {NoStop}%
\bibitem [{\citenamefont {Vilenkin}(1994)}]{Vi94}%
  \BibitemOpen
  \bibfield  {author} {\bibinfo {author} {\bibfnamefont {A.}~\bibnamefont
  {Vilenkin}},\ }\bibfield  {title} {\bibinfo {title} {Approaches to quantum
  cosmology},\ }\href@noop {} {\bibfield  {journal} {\bibinfo  {journal} {Phys.
  Rev. D}\ }\textbf {\bibinfo {volume} {50}},\ \bibinfo {pages} {2581}
  (\bibinfo {year} {1994})}\BibitemShut {NoStop}%
\bibitem [{\citenamefont {Atkatz}(1994)}]{At94}%
  \BibitemOpen
  \bibfield  {author} {\bibinfo {author} {\bibfnamefont {D.}~\bibnamefont
  {Atkatz}},\ }\bibfield  {title} {\bibinfo {title} {Quantum cosmology for
  pedestrians},\ }\href@noop {} {\bibfield  {journal} {\bibinfo  {journal} {Am.
  J. Phys.}\ }\textbf {\bibinfo {volume} {62}},\ \bibinfo {pages} {619}
  (\bibinfo {year} {1994})}\BibitemShut {NoStop}%
\bibitem [{\citenamefont {Linde}(2008)}]{LI08}%
  \BibitemOpen
  \bibfield  {author} {\bibinfo {author} {\bibfnamefont {A.}~\bibnamefont
  {Linde}},\ }\bibinfo {title} {Inflationary cosmology},\ in\ \href@noop {}
  {\emph {\bibinfo {booktitle} {Inflationary Cosmology}}},\ \bibinfo {series}
  {Lecture Notes in Physics}, Vol.\ \bibinfo {volume} {738}\ (\bibinfo
  {publisher} {Springer},\ \bibinfo {year} {2008})\ pp.\ \bibinfo {pages}
  {1--54}\BibitemShut {NoStop}%
\bibitem [{\citenamefont {Handley}\ \emph {et~al.}(2014)\citenamefont
  {Handley}, \citenamefont {Brechet}, \citenamefont {Lasenby},\ and\
  \citenamefont {Hobson}}]{HAN14}%
  \BibitemOpen
  \bibfield  {author} {\bibinfo {author} {\bibfnamefont {W.}~\bibnamefont
  {Handley}}, \bibinfo {author} {\bibfnamefont {S.}~\bibnamefont {Brechet}},
  \bibinfo {author} {\bibfnamefont {A.}~\bibnamefont {Lasenby}},\ and\ \bibinfo
  {author} {\bibfnamefont {M.~P.}\ \bibnamefont {Hobson}},\ }\bibfield  {title}
  {\bibinfo {title} {Kinetic initial conditions for inflation},\ }\href@noop {}
  {\bibfield  {journal} {\bibinfo  {journal} {Phys. Rev. D}\ }\textbf {\bibinfo
  {volume} {89}},\ \bibinfo {pages} {063505} (\bibinfo {year}
  {2014})}\BibitemShut {NoStop}%
\bibitem [{\citenamefont {Handley}\ \emph {et~al.}(2019)\citenamefont
  {Handley}, \citenamefont {Lasenby},\ and\ \citenamefont {Hobson}}]{HAN19}%
  \BibitemOpen
  \bibfield  {author} {\bibinfo {author} {\bibfnamefont {W.}~\bibnamefont
  {Handley}}, \bibinfo {author} {\bibfnamefont {A.}~\bibnamefont {Lasenby}},\
  and\ \bibinfo {author} {\bibfnamefont {M.}~\bibnamefont {Hobson}},\
  }\bibfield  {title} {\bibinfo {title} {Kinetically dominated curved
  universes: Logolinear series expansions},\ }\href@noop {} {\bibfield
  {journal} {\bibinfo  {journal} {Phys. Rev. D}\ }\textbf {\bibinfo {volume}
  {99}},\ \bibinfo {pages} {123512} (\bibinfo {year} {2019})}\BibitemShut
  {NoStop}%
\bibitem [{\citenamefont {Haddadin}\ and\ \citenamefont
  {Handley}(2018)}]{HAD18}%
  \BibitemOpen
  \bibfield  {author} {\bibinfo {author} {\bibfnamefont {W.~I.~J.}\
  \bibnamefont {Haddadin}}\ and\ \bibinfo {author} {\bibfnamefont {W.~J.}\
  \bibnamefont {Handley}},\ }\bibfield  {title} {\bibinfo {title} {Rapid
  numerical solutions for the {M}ukhanov-{S}azaki equation},\ }\href@noop {}
  {\bibfield  {journal} {\bibinfo  {journal} {arXiv:1809.11095}\ } (\bibinfo
  {year} {2018})}\BibitemShut {NoStop}%
\bibitem [{\citenamefont {Hergt}\ \emph {et~al.}(2019)\citenamefont {Hergt},
  \citenamefont {Handley}, \citenamefont {Hobson},\ and\ \citenamefont
  {Lasenby}}]{HER19}%
  \BibitemOpen
  \bibfield  {author} {\bibinfo {author} {\bibfnamefont {L.}~\bibnamefont
  {Hergt}}, \bibinfo {author} {\bibfnamefont {W.}~\bibnamefont {Handley}},
  \bibinfo {author} {\bibfnamefont {M.}~\bibnamefont {Hobson}},\ and\ \bibinfo
  {author} {\bibfnamefont {A.}~\bibnamefont {Lasenby}},\ }\bibfield  {title}
  {\bibinfo {title} {Constraining the kinetically dominated universe},\
  }\href@noop {} {\bibfield  {journal} {\bibinfo  {journal} {Phys. Rev. D}\
  }\textbf {\bibinfo {volume} {100}},\ \bibinfo {pages} {023501} (\bibinfo
  {year} {2019})}\BibitemShut {NoStop}%
\bibitem [{\citenamefont {Medina}\ and\ \citenamefont {{Mart\'{\i}nez
  Alonso}}(2020)}]{ME20}%
  \BibitemOpen
  \bibfield  {author} {\bibinfo {author} {\bibfnamefont {E.}~\bibnamefont
  {Medina}}\ and\ \bibinfo {author} {\bibfnamefont {L.}~\bibnamefont
  {{Mart\'{\i}nez Alonso}}},\ }\bibfield  {title} {\bibinfo {title} {Kinetic
  dominance and psi series in the {H}amilton-{J}acobi formulation of inflaton
  models},\ }\href@noop {} {\bibfield  {journal} {\bibinfo  {journal} {Phys.
  Rev. D}\ }\textbf {\bibinfo {volume} {102}},\ \bibinfo {pages} {103517}
  (\bibinfo {year} {2020})}\BibitemShut {NoStop}%
\bibitem [{\citenamefont {Steinhardt}\ and\ \citenamefont
  {Turner}(1984)}]{STEIN84}%
  \BibitemOpen
  \bibfield  {author} {\bibinfo {author} {\bibfnamefont {P.~J.}\ \bibnamefont
  {Steinhardt}}\ and\ \bibinfo {author} {\bibfnamefont {M.~S.}\ \bibnamefont
  {Turner}},\ }\bibfield  {title} {\bibinfo {title} {Prescription for
  successful new inflation},\ }\href@noop {} {\bibfield  {journal} {\bibinfo
  {journal} {Phys. Rev. D}\ }\textbf {\bibinfo {volume} {29}},\ \bibinfo
  {pages} {2162} (\bibinfo {year} {1984})}\BibitemShut {NoStop}%
\bibitem [{\citenamefont {Stewart}\ and\ \citenamefont {Lyth}(1993)}]{STE93}%
  \BibitemOpen
  \bibfield  {author} {\bibinfo {author} {\bibfnamefont {E.~D.}\ \bibnamefont
  {Stewart}}\ and\ \bibinfo {author} {\bibfnamefont {D.~H.}\ \bibnamefont
  {Lyth}},\ }\bibfield  {title} {\bibinfo {title} {A more accurate analytic
  calculation of the spectrum of cosmological perturbations produced during
  inflation},\ }\href@noop {} {\bibfield  {journal} {\bibinfo  {journal} {Phys.
  Lett. B}\ }\textbf {\bibinfo {volume} {302}},\ \bibinfo {pages} {171}
  (\bibinfo {year} {1993})}\BibitemShut {NoStop}%
\bibitem [{\citenamefont {Liddle}\ \emph {et~al.}(1994)\citenamefont {Liddle},
  \citenamefont {Parsons},\ and\ \citenamefont {Barrow}}]{LID94}%
  \BibitemOpen
  \bibfield  {author} {\bibinfo {author} {\bibfnamefont {A.~R.}\ \bibnamefont
  {Liddle}}, \bibinfo {author} {\bibfnamefont {P.}~\bibnamefont {Parsons}},\
  and\ \bibinfo {author} {\bibfnamefont {J.~D.}\ \bibnamefont {Barrow}},\
  }\bibfield  {title} {\bibinfo {title} {Formalizing the slow-roll
  approximation in inflation},\ }\href@noop {} {\bibfield  {journal} {\bibinfo
  {journal} {Phys. Rev. D}\ }\textbf {\bibinfo {volume} {50}},\ \bibinfo
  {pages} {7222} (\bibinfo {year} {1994})}\BibitemShut {NoStop}%
\bibitem [{\citenamefont {Lidsey}\ \emph {et~al.}(1997)\citenamefont {Lidsey},
  \citenamefont {Liddle}, \citenamefont {Kolb}, \citenamefont {Copeland},
  \citenamefont {Barreiro},\ and\ \citenamefont {Abney}}]{LIDS97}%
  \BibitemOpen
  \bibfield  {author} {\bibinfo {author} {\bibfnamefont {J.~E.}\ \bibnamefont
  {Lidsey}}, \bibinfo {author} {\bibfnamefont {A.~R.}\ \bibnamefont {Liddle}},
  \bibinfo {author} {\bibfnamefont {E.~W.}\ \bibnamefont {Kolb}}, \bibinfo
  {author} {\bibfnamefont {E.~J.}\ \bibnamefont {Copeland}}, \bibinfo {author}
  {\bibfnamefont {T.}~\bibnamefont {Barreiro}},\ and\ \bibinfo {author}
  {\bibfnamefont {M.}~\bibnamefont {Abney}},\ }\bibfield  {title} {\bibinfo
  {title} {Reconstructing the inflaton potential---an overview},\ }\href@noop
  {} {\bibfield  {journal} {\bibinfo  {journal} {Rev. Mod. Phys.}\ }\textbf
  {\bibinfo {volume} {69}},\ \bibinfo {pages} {373} (\bibinfo {year}
  {1997})}\BibitemShut {NoStop}%
\bibitem [{\citenamefont {Bassett}\ \emph {et~al.}(2006)\citenamefont
  {Bassett}, \citenamefont {Tsujikawa},\ and\ \citenamefont {Wands}}]{BA06}%
  \BibitemOpen
  \bibfield  {author} {\bibinfo {author} {\bibfnamefont {B.~A.}\ \bibnamefont
  {Bassett}}, \bibinfo {author} {\bibfnamefont {S.}~\bibnamefont {Tsujikawa}},\
  and\ \bibinfo {author} {\bibfnamefont {D.}~\bibnamefont {Wands}},\ }\bibfield
   {title} {\bibinfo {title} {Inflation dynamics and reheating},\ }\href@noop
  {} {\bibfield  {journal} {\bibinfo  {journal} {Rev. Mod. Phys.}\ }\textbf
  {\bibinfo {volume} {78}},\ \bibinfo {pages} {537} (\bibinfo {year}
  {2006})}\BibitemShut {NoStop}%
\bibitem [{\citenamefont {Weinberg}(2008)}]{WE08}%
  \BibitemOpen
  \bibfield  {author} {\bibinfo {author} {\bibfnamefont {S.}~\bibnamefont
  {Weinberg}},\ }\href@noop {} {\emph {\bibinfo {title} {Cosmology}}}\
  (\bibinfo  {publisher} {Oxford University Press},\ \bibinfo {year}
  {2008})\BibitemShut {NoStop}%
\bibitem [{\citenamefont {Boyanovsky}\ \emph {et~al.}(2009)\citenamefont
  {Boyanovsky}, \citenamefont {Destri}, \citenamefont {de~Vega},\ and\
  \citenamefont {Sanchez}}]{BO09}%
  \BibitemOpen
  \bibfield  {author} {\bibinfo {author} {\bibfnamefont {D.}~\bibnamefont
  {Boyanovsky}}, \bibinfo {author} {\bibfnamefont {C.}~\bibnamefont {Destri}},
  \bibinfo {author} {\bibfnamefont {H.}~\bibnamefont {de~Vega}},\ and\ \bibinfo
  {author} {\bibfnamefont {N.}~\bibnamefont {Sanchez}},\ }\bibfield  {title}
  {\bibinfo {title} {The effective theory of inflation in the standard model of
  the universe and the {CMB+LSS} data analysis},\ }\href@noop {} {\bibfield
  {journal} {\bibinfo  {journal} {Int.J. Mod. Phys. A}\ }\textbf {\bibinfo
  {volume} {24}},\ \bibinfo {pages} {3669} (\bibinfo {year}
  {2009})}\BibitemShut {NoStop}%
\bibitem [{\citenamefont {Lasenby}\ and\ \citenamefont {Doran}(2005)}]{LA05}%
  \BibitemOpen
  \bibfield  {author} {\bibinfo {author} {\bibfnamefont {A.}~\bibnamefont
  {Lasenby}}\ and\ \bibinfo {author} {\bibfnamefont {C.}~\bibnamefont
  {Doran}},\ }\bibfield  {title} {\bibinfo {title} {Closed universes, de
  {S}itter space, and inflation},\ }\href@noop {} {\bibfield  {journal}
  {\bibinfo  {journal} {Phys. Rev. D}\ }\textbf {\bibinfo {volume} {71}},\
  \bibinfo {pages} {063502} (\bibinfo {year} {2005})}\BibitemShut {NoStop}%
\bibitem [{\citenamefont {Destri}\ \emph {et~al.}(2010)\citenamefont {Destri},
  \citenamefont {de~Vega},\ and\ \citenamefont {Sanchez}}]{DE10}%
  \BibitemOpen
  \bibfield  {author} {\bibinfo {author} {\bibfnamefont {C.}~\bibnamefont
  {Destri}}, \bibinfo {author} {\bibfnamefont {H.}~\bibnamefont {de~Vega}},\
  and\ \bibinfo {author} {\bibfnamefont {N.~G.}\ \bibnamefont {Sanchez}},\
  }\bibfield  {title} {\bibinfo {title} {The pre-inflationary and inflationary
  fast-roll eras and their signatures in the low {CMB} multipoles},\
  }\href@noop {} {\bibfield  {journal} {\bibinfo  {journal} {Phys. Rev. D}\
  }\textbf {\bibinfo {volume} {81}},\ \bibinfo {pages} {063520} (\bibinfo
  {year} {2010})}\BibitemShut {NoStop}%
\bibitem [{\citenamefont {DeWitt}(1967)}]{DE67}%
  \BibitemOpen
  \bibfield  {author} {\bibinfo {author} {\bibfnamefont {B.~S.}\ \bibnamefont
  {DeWitt}},\ }\bibfield  {title} {\bibinfo {title} {Quantum theory of gravity.
  {I}. {T}he canonical theory},\ }\href@noop {} {\bibfield  {journal} {\bibinfo
   {journal} {Phys. Rev.}\ }\textbf {\bibinfo {volume} {160}},\ \bibinfo
  {pages} {1113} (\bibinfo {year} {1967})}\BibitemShut {NoStop}%
\bibitem [{\citenamefont {Hawking}\ and\ \citenamefont {Page}(1986)}]{HP86}%
  \BibitemOpen
  \bibfield  {author} {\bibinfo {author} {\bibfnamefont {S.~W.}\ \bibnamefont
  {Hawking}}\ and\ \bibinfo {author} {\bibfnamefont {D.~N.}\ \bibnamefont
  {Page}},\ }\bibfield  {title} {\bibinfo {title} {Operator ordering and the
  flatness of the universe},\ }\href@noop {} {\bibfield  {journal} {\bibinfo
  {journal} {Nuc. Phys. B}\ }\textbf {\bibinfo {volume} {264}},\ \bibinfo
  {pages} {185} (\bibinfo {year} {1986})}\BibitemShut {NoStop}%
\bibitem [{\citenamefont {Halliwell}(1991)}]{ha09}%
  \BibitemOpen
  \bibfield  {author} {\bibinfo {author} {\bibfnamefont {J.}~\bibnamefont
  {Halliwell}},\ }\bibfield  {title} {\bibinfo {title} {{Introductory Lectures
  on Quantum Cosmology}},\ }in\ \href@noop {} {\emph {\bibinfo {booktitle}
  {{Jerusalem Winter School on Quantum Cosmology and Baby Universes}}}},\
  \bibinfo {editor} {edited by\ \bibinfo {editor} {\bibfnamefont {P.~T.}\
  \bibnamefont {Coleman~S.}, \bibfnamefont {Hartle J.~B.}}\ and\ \bibinfo
  {editor} {\bibfnamefont {S.}~\bibnamefont {Weinberg}}}\ (\bibinfo
  {publisher} {World Scientific},\ \bibinfo {address} {Singapore},\ \bibinfo
  {year} {1991})\BibitemShut {NoStop}%
\bibitem [{\citenamefont {Felder}\ \emph {et~al.}(2002)\citenamefont {Felder},
  \citenamefont {Frolov}, \citenamefont {Kofman},\ and\ \citenamefont
  {Linde}}]{FE02}%
  \BibitemOpen
  \bibfield  {author} {\bibinfo {author} {\bibfnamefont {G.}~\bibnamefont
  {Felder}}, \bibinfo {author} {\bibfnamefont {A.}~\bibnamefont {Frolov}},
  \bibinfo {author} {\bibfnamefont {L.}~\bibnamefont {Kofman}},\ and\ \bibinfo
  {author} {\bibfnamefont {A.}~\bibnamefont {Linde}},\ }\bibfield  {title}
  {\bibinfo {title} {Cosmology with negative potentials},\ }\href@noop {}
  {\bibfield  {journal} {\bibinfo  {journal} {Phys. Rev. D}\ }\textbf {\bibinfo
  {volume} {66}},\ \bibinfo {pages} {023507} (\bibinfo {year}
  {2002})}\BibitemShut {NoStop}%
\bibitem [{\citenamefont {Berry}\ and\ \citenamefont {Mount}(1972)}]{BM72}%
  \BibitemOpen
  \bibfield  {author} {\bibinfo {author} {\bibfnamefont {M.~V.}\ \bibnamefont
  {Berry}}\ and\ \bibinfo {author} {\bibfnamefont {K.~E.}\ \bibnamefont
  {Mount}},\ }\bibfield  {title} {\bibinfo {title} {Semiclassical
  approximations in wave mechanics},\ }\href@noop {} {\bibfield  {journal}
  {\bibinfo  {journal} {Rep. Prog. Phys.}\ }\textbf {\bibinfo {volume} {35}},\
  \bibinfo {pages} {315} (\bibinfo {year} {1972})}\BibitemShut {NoStop}%
\bibitem [{\citenamefont {Galindo}\ and\ \citenamefont {Pascual}(1991)}]{GA91}%
  \BibitemOpen
  \bibfield  {author} {\bibinfo {author} {\bibfnamefont {A.}~\bibnamefont
  {Galindo}}\ and\ \bibinfo {author} {\bibfnamefont {P.}~\bibnamefont
  {Pascual}},\ }\href@noop {} {\emph {\bibinfo {title} {Quantum {M}echanics
  II}}}\ (\bibinfo  {publisher} {Springer},\ \bibinfo {year}
  {1991})\BibitemShut {NoStop}%
\bibitem [{\citenamefont {Silverstone}(1985)}]{SI85}%
  \BibitemOpen
  \bibfield  {author} {\bibinfo {author} {\bibfnamefont {H.~J.}\ \bibnamefont
  {Silverstone}},\ }\bibfield  {title} {\bibinfo {title} {{JWKB}
  connection-formula problem revisited via {Borel} summation},\ }\href@noop {}
  {\bibfield  {journal} {\bibinfo  {journal} {Phys. Rev. Lett.}\ }\textbf
  {\bibinfo {volume} {55}},\ \bibinfo {pages} {2523} (\bibinfo {year}
  {1985})}\BibitemShut {NoStop}%
\bibitem [{\citenamefont {Langer}(1937)}]{LA37}%
  \BibitemOpen
  \bibfield  {author} {\bibinfo {author} {\bibfnamefont {L.~E.}\ \bibnamefont
  {Langer}},\ }\bibfield  {title} {\bibinfo {title} {{On the Connection
  Formulas and the Solutions of the Wave Equation}},\ }\href@noop {} {\bibfield
   {journal} {\bibinfo  {journal} {Phys. Rev.}\ }\textbf {\bibinfo {volume}
  {51}},\ \bibinfo {pages} {669} (\bibinfo {year} {1937})}\BibitemShut
  {NoStop}%
\bibitem [{\citenamefont {Cherry}(1950)}]{CH50}%
  \BibitemOpen
  \bibfield  {author} {\bibinfo {author} {\bibfnamefont {T.~M.}\ \bibnamefont
  {Cherry}},\ }\bibfield  {title} {\bibinfo {title} {Uniform asymptotic
  formulae for function with transition points},\ }\href@noop {} {\bibfield
  {journal} {\bibinfo  {journal} {Trans. Am. Math. Soc.}\ }\textbf {\bibinfo
  {volume} {68}},\ \bibinfo {pages} {224} (\bibinfo {year} {1950})}\BibitemShut
  {NoStop}%
\bibitem [{\citenamefont {Shen}\ and\ \citenamefont
  {Silverstone}(2004)}]{SS04}%
  \BibitemOpen
  \bibfield  {author} {\bibinfo {author} {\bibfnamefont {H.}~\bibnamefont
  {Shen}}\ and\ \bibinfo {author} {\bibfnamefont {H.~J.}\ \bibnamefont
  {Silverstone}},\ }\bibfield  {title} {\bibinfo {title} {{JWKB} method as an
  exact technique},\ }\href@noop {} {\bibfield  {journal} {\bibinfo  {journal}
  {Int. J. Quantum Chem.}\ }\textbf {\bibinfo {volume} {99}},\ \bibinfo {pages}
  {336} (\bibinfo {year} {2004})}\BibitemShut {NoStop}%
\bibitem [{\citenamefont {Barvinsky}\ \emph {et~al.}(2015)\citenamefont
  {Barvinsky}, \citenamefont {Kamenshchik},\ and\ \citenamefont
  {Nesterov}}]{BAR15}%
  \BibitemOpen
  \bibfield  {author} {\bibinfo {author} {\bibfnamefont {A.~O.}\ \bibnamefont
  {Barvinsky}}, \bibinfo {author} {\bibfnamefont {A.~Y.}\ \bibnamefont
  {Kamenshchik}},\ and\ \bibinfo {author} {\bibfnamefont {D.}~\bibnamefont
  {Nesterov}},\ }\bibfield  {title} {\bibinfo {title} {Origin of inflation in
  {CFT} driven cosmology: ${R}^2$-gravity and non-minimally coupled inflaton
  models},\ }\href@noop {} {\bibfield  {journal} {\bibinfo  {journal} {Eur.
  Phys. J. C}\ }\textbf {\bibinfo {volume} {75}},\ \bibinfo {pages} {1}
  (\bibinfo {year} {2015})}\BibitemShut {NoStop}%
\bibitem [{\citenamefont {Barvinsky}\ \emph {et~al.}(2016)\citenamefont
  {Barvinsky}, \citenamefont {Kamenshchik},\ and\ \citenamefont
  {Nesterov}}]{BAR16}%
  \BibitemOpen
  \bibfield  {author} {\bibinfo {author} {\bibfnamefont {A.~O.}\ \bibnamefont
  {Barvinsky}}, \bibinfo {author} {\bibfnamefont {A.~Y.}\ \bibnamefont
  {Kamenshchik}},\ and\ \bibinfo {author} {\bibfnamefont {D.~V.}\ \bibnamefont
  {Nesterov}},\ }\bibfield  {title} {\bibinfo {title} {New type of hill-top
  inflation},\ }\href@noop {} {\bibfield  {journal} {\bibinfo  {journal} {J.
  Cosmol. Astropart. Phys.}\ }\textbf {\bibinfo {volume} {2016}}\bibinfo
  {number} { (01)},\ \bibinfo {pages} {036}}\BibitemShut {NoStop}%
\bibitem [{\citenamefont {Abramowitz}\ and\ \citenamefont
  {Stegun}(1967)}]{AS67}%
  \BibitemOpen
\bibfield  {number} {  }\bibinfo {editor} {\bibfnamefont {M.}~\bibnamefont
  {Abramowitz}}\ and\ \bibinfo {editor} {\bibfnamefont {I.~A.}\ \bibnamefont
  {Stegun}},\ eds.,\ \href@noop {} {\emph {\bibinfo {title} {Handbook of
  mathematical functions}}}\ (\bibinfo  {publisher} {National Bureau of
  Standards},\ \bibinfo {year} {1967})\BibitemShut {NoStop}%
\end{thebibliography}%
%%%%%%%%%%%%%%%%%%%%%%%%%%%%%%%%%%%%%%%%%%%%%%%%%%%%%%%%%%%%%%%%%
\end{document}